\documentclass[fleqn,usenatbib]{mnras}

\DeclareRobustCommand{\VAN}[3]{#2}
\let\VANthebibliography\thebibliography
\def\thebibliography{\DeclareRobustCommand{\VAN}[3]{##3}\VANthebibliography}

\usepackage{graphicx}
\usepackage{amsmath}
\usepackage{amssymb}
\usepackage{pdflscape}
\usepackage{longtable}
\usepackage{booktabs}
\usepackage{natbib}
\usepackage{enumitem}
\usepackage{booktabs}
\usepackage{amsmath}
\usepackage{array}
\usepackage{wasysym}
\newcommand{\msun}{M$_\odot$}

\newcommand{\mbh}{M$_\bullet$}
\newcommand{\mmbh}{M_\bullet}

\newcommand{\Htwo}{H$_{2}$}

\newcommand{\Fe}{[\ion{Fe}{ii}]}

\newcommand{\OIII}{[\ion{O}{iii}]}
\newcommand{\OIIIL}{[\ion{O}{iii}]$\lambda5007$}
\newcommand{\NII}{[\ion{N}{ii}]}

\newcommand{\Ha}{H\,$\alpha${}}
\newcommand{\Hb}{H\,$\beta${}}
\newcommand{\Hg}{H\,$\gamma${}}

\newcommand{\pab}{Pa\,$\beta${}}
\newcommand{\paa}{Pa\,$\alpha${}}
\newcommand{\pag}{Pa\,$\gamma${}}
\newcommand{\pad}{Pa\,$\delta${}}
\newcommand{\brg}{Br\,$\gamma${}}

\newcommand{\ecs}{erg~cm\pwr{-2}~s\pwr{-1}}
\newcommand{\es}{erg~s\pwr{-1}}

\newcommand{\LBCont}{$L_{5100}$}
\newcommand{\kms}{km~s\pwr{-1}}

\newcommand{\ebv}{$E(B-V)$}
\newcommand{\ebvsub}{$E_{B-V}$}

\newcommand{\pwr}[1]{$^{#1}$}

\makeatletter
\newcommand*{\rom}[1]{\expandafter\@slowromancap\romannumeral #1@}
\makeatother
\makeatletter
\def\blfootnote{\xdef\@thefnmark{}\@footnotetext}
\makeatother

\newcolumntype{R}[1]{>{\RaggedLeft\arraybackslash}p{#1}}
\newcolumntype{C}[1]{>{\centering\arraybackslash}p{#1}}

\title[Narrow-Line Seyfert 1 IR Survey]{The Southern Hemisphere Narrow-Line Seyfert 1 Infrared Survey}

\author[A.M. Durr\'{e} et al.]{Mark Durr\'{e}$^{1}$\thanks{E-mail: mdurre@swin.edu.au}
and Jeremy Mould$^{1}$
\\
$^{1}$Centre for Astrophysics \& Supercomputing, Swinburne University, Hawthorn VIC 3122, Australia\\
}

\date{Accepted XXX. Received YYY; in original form ZZZ}

\pubyear{2021}

\begin{document}
\label{firstpage}
\pagerange{\pageref{firstpage}--\pageref{lastpage}}
\maketitle
\defcitealias{Chen2018}{C18}
\defcitealias{Schmidt2018}{S18}

\begin{abstract}
We present a near-infrared spectroscopic survey of narrow-line Seyfert 1 galaxies in the southern hemisphere (using the SOFI instrument on the ESO-NTT telescope), sampled from optical surveys. We examine the kinematics of the broad-line region, probed by the emission line width of hydrogen (\paa{} and \Hb). We observed 57 objects, of which we could firmly measure \paa{} in 49 cases. We find that a single Lorentzian fit (preferred on theoretical grounds) is preferred over multi-component Gaussian fits to the line profiles; a lack of narrow-line region emission, overwhelmed by the pole-on view of the broad line region (BLR) light, supports this. We recompute the catalog black hole (BH) mass estimates, using the values of FWHM and luminosity of \Hb{}, both from catalog values and re-fitted Lorentzian values. We find a relationship slope greater than unity compared to the catalog values. We ascribe this to contamination by galactic light or difficulties with line flux measurements. However, the comparison of masses computed by the fitted Lorentzian and Gaussian measurements show a slope close to unity. Comparing the BH masses estimated from both \paa{}and \Hb, the line widths and fluxes shows deviations from expected; in general, however, the computed BH masses are comparable. We posit a scenario where an intermixture of dusty and dust-free clouds (or alternately a structured atmosphere) differentially absorbs the line radiation of the BLR, due to dust absorption and hydrogen bound-free absorption.
\end{abstract}




\section{Introduction}
Narrow-line Seyfert 1 (NLSy1) galaxies are a class of active galactic nuclei (AGN) where the broad-line region (BLR) \Hb{} emission lines show characteristic velocities full-width half-maximum (FWHM) $<$ 2000 \kms{} and flux ratio of \OIII/\Hb $<$ 3, as well as prominent Fe II lines \citep{Osterbrock1985,Goodrich1989}. These characteristics can be interpreted as either (a) a relatively under-massive black hole (BH) typically 10\pwr{6-8} \msun{}(vs. 10\pwr{7-9} \msun{} for typical broad-line Seyfert 1s), with correspondingly high Eddington ratio/accretion rate (0.1 to 1), suggesting a young, fast-growing phase of the AGN \citep{Viswanath2019a,Berton2015}, or (b) the BLR is disk-like and viewed pole on; thus Doppler broadening is suppressed and the apparent low mass of the BH is merely an inclination effect \citep{Jolley2008}. The BLR models of \cite{Goad2012}, which have a bowl-shaped geometry combined with turbulence, naturally favour the second view.

Radio-loud NLSy1 galaxies (only about 7\% of all NLSy1 galaxies) share some of the characteristics of blazars, including $\gamma$-ray emission \citep{Abdo2009} and radio beamed jets \citep{Foschini2015}, albeit at lower luminosity and BH mass. The similarity to blazars suggest a pole-on orientation \citep{Decarli2008,Decarli2011}; this corrects for the lower BH masses and high Eddington ratios, which are then more in line with their  host galaxy scaling relationships, e.g. $\mmbh-\sigma_{*}$ \cite[e.g.][]{Ryan2007}.

We present results from an infrared spectroscopic survey of this class of objects. Our primary objective is to study the kinematics of the gas in the line emitting region, which are probed by the \paa{} and \Hb{} emission lines.  Observations in the near infra-red (NIR) have the advantage of penetrating obscuring dust; the \paa{} emission line flux (1875 nm) is less affected than the \Hb{} flux (486.3 nm) and thus sees ``deeper''.

The mass of the BH can be estimated from the luminosity and full-width half-maximum (FWHM) velocity of the broad hydrogen emission lines from the BLR \citep[see e.g.][and references therein]{Kim2010,Popovic2020}. We will compare methods of estimating this mass with different emission lines, line profile models and continuum parameters.

This paper is organized as follows. In Section \ref{sec:Sec2}, we present the sample selection; in Section \ref{sec:Sec3}, we discuss the observations, data reduction and emission-line measurement methods and results and in Section \ref{sec:Sec4}, we estimating the black hole masses by the different methods (including comparison with optical emission lines) and calculate the extinctions in the BLR. In Section \ref{sec:Sec5}, we discuss our results, and in Section \ref{sec:Sec6}, we present our conclusions.

For consistency with the source surveys for our sample, we adopt the cosmological parameters $H_0 = 70$ km{} s$^{-1}$, $\Omega_{m} = 0.27$ and $\Omega_{\Lambda} = 0.73$.

\section{Sample Selection}
\label{sec:Sec2}
Our selection of galaxies are from two previous surveys. \cite{Chen2018}\footnote{\url{http://cdsarc.u-strasbg.fr/viz-bin/qcat?J/A+A/615/A167}} (hereafter \citetalias{Chen2018}) classified 167 southern hemisphere galaxies as NLSy1s from the Six-degree Field Galaxy Survey (6dFGS) final data release (DR3) \citep{Jones2004,Jones2009}\footnote{\url{http://www-wfau.roe.ac.uk/6dFGS/}}; they derived flux-calibrated spectra and found strong correlations between the monochromatic luminosity at 5100\AA{} and the luminosities of \Hb{} and \OIIIL{} lines, with typical SMBH masses (10\pwr{6}--10\pwr{8} \msun) and high Eddington ratios (a median of 0.96) for the AGN class. The survey of \cite{Schmidt2018}\footnote{\url{http://cdsarc.u-strasbg.fr/viz-bin/qcat?J/A+A/615/A13}} (hereafter \citetalias{Schmidt2018}) (28 galaxies) was focused on \OIIIL{} line profile asymmetries associated with outflows, confirming the correlation between the blueshift and FWHM of the line core, between the outflow velocity and the black hole mass and with the width of the broad component of \Hb. Removing objects common to the two catalogs gave a target list of 189 galaxies with declination $<10\degr$. The redshifts ($z$) are in the range $0.01<z<0.5$ (which corresponds to a luminosity distance range of 44 to 2855 Mpc).

\begin{figure*}
\centering
\includegraphics[width=.8\linewidth]{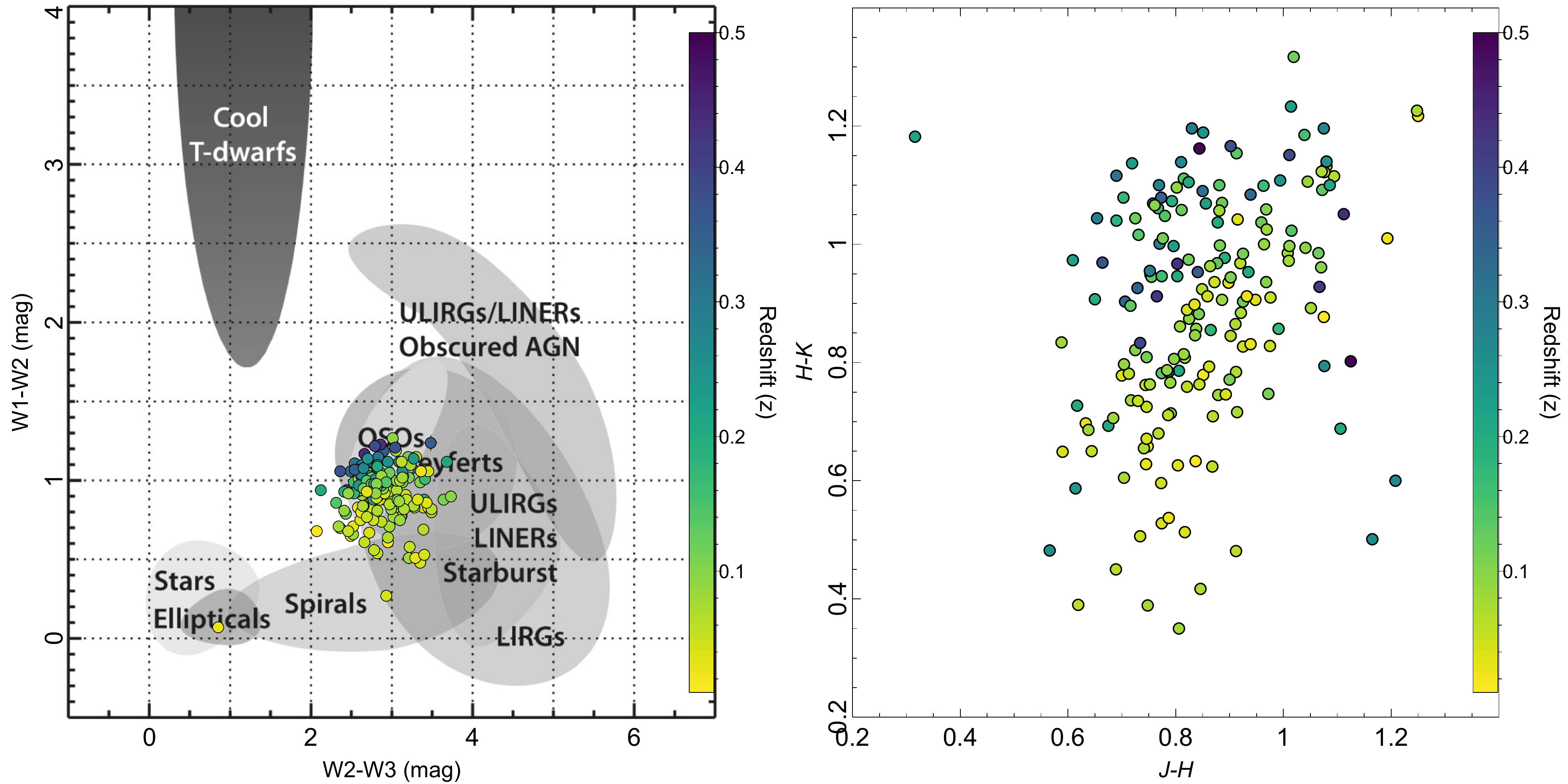}
\caption{(Left panel) \textit{WISE} $W1-W2$ vs. $W2-W3$ color-color plot for NLSy1 sample, with point colors indicating redshift; note the trend of locus with redshift. (Right panel) 2MASS $J-H$ vs. $H-K$ color-color plot for NLSy1 sample, with point colors indicating redshift, showing a similar trend of locus with redshift.}
\label{fig:shirnlsresults01}
\end{figure*}

\begin{figure}
\centering
\includegraphics[width=.7\linewidth]{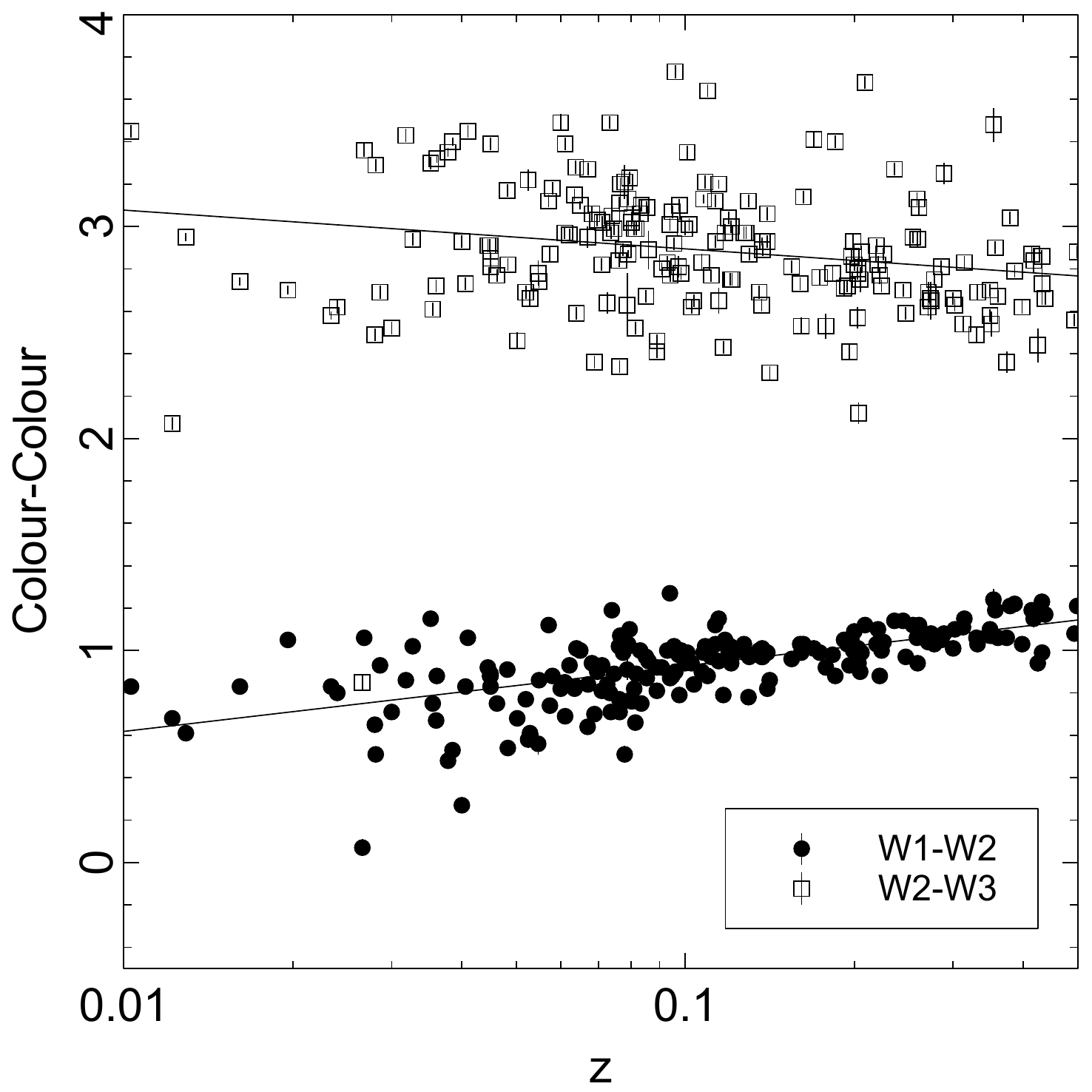}
\caption{\textit{WISE} color trend with redshift for the sample, with linear fits.}
\label{fig:shirnlsresults02}
\end{figure}

Hydrogen emission line flux can come from both AGN activity and star formation (SF). To determine whether SF is significant in our sample, we use mid infrared (MIR) color-color plots from \textit{WISE} satellite data. We retrieved the \textit{WISE} magnitudes from the AllWISE catalog \citep{Wright2010}; Figure \ref{fig:shirnlsresults01} (left panel) shows the \textit{W2-W3} vs. \textit{W1-W2} plot \citep[overlaying the color-color diagram from][their Figure 12]{Wright2010}, with point colors indicating the object redshift. Most galaxies are in the Seyfert/QSO locus, with some having an admixture of stellar light. None are within the starburst (SF) region, which indicates minimal, if any, star formation in our sample. The single point near (0.5,0) is due to sky contamination by galactic cirrus. We note that the plot exhibits a trend with redshift; this is expected as the spectral peak of the warm dust (with maximum in the \textit{W2} band) is red-shifted, effectively a K-correction. Figure \ref{fig:shirnlsresults02} shows the trend of colors with redshift. In a similar manner, we also plotted the 2MASS \textit{JHK} color-color diagram (Figure \ref{fig:shirnlsresults01} - right panel); this also shows a trend with redshift as per the \textit{WISE} diagram.

We also test the reliability of \textit{WISE} color selection on our sample; we use the 90\% reliability criterion (R90) from \cite{Assef2018}, finding that our objects are selected as AGN at this reliability in 93.5\% (177/189) of cases. Those that are rejected have a redshift population significantly less than those accepted; the Student's one-tailed t-test on the redshifts shows a p-value of $6.86\times10^{-4}$, i.e. we can reject the hypothesis that the populations are the same (rejected - N=12, $\bar{z}=0.045, \sigma_z=0.018$; selected - N=177, $\bar{z}=0.152, \sigma_z=0.1136$). This is most likely due to contamination by galactic stellar light.

\section{Methods}
\label{sec:Sec3}
\subsection{Observations and Data Reduction}
\label{sec:observations}
We observed a sub-sample of these galaxies with the SofI (Son of ISAAC) near infrared (NIR) spectrometer \citep{Moorwood1998} at the Nasmyth A focus of the ESO 3.5-m New Technology Telescope (NTT) at the European Southern Observatory (ESO) in La Silla (Chile), in the program 0103.B-0504(B) (observation dates 25--27 August 2019). To maximize efficiency, our observing strategy was to prioritize objects based on a figure of merit, which selected both on 2MASS \textit{Ks} magnitude \citep[with sample range of 10.9--17.4 mag]{Skrutskie2006} and on total \Hb{} flux (from $1.4-590 \times 10^{-15}$ \es) as taken from the \citetalias{Chen2018} and \citetalias{Schmidt2018} catalogs, i.e. we selected both bright galaxies and strong emission lines.

The GRF (red) grism was used with a 1\arcsec{} slit width, for a spectral range of 1495\nobreakdash-2518 nm and a spectral resolution $R\sim600$. For the redshift range, this allows for observation of at least one hydrogen line (\paa, \brg{} or \pab) plus \Fe$\lambda1644$ (all objects) or \Fe$\lambda1257$ (56/189) plus \Htwo{} (133/189). The \paa{} emission line falls within the atmospheric absorption band ($\sim1800-1950$ nm) in 12 of 171 objects where the line was within the spectral range of the instrument, but even then the \paa{} line is strong enough to be measured in some of these cases. 

Over the three nights, 55 objects were successfully observed in the standard ABBA spectroscopic mode (the observing template "SOFI\_spec\_obs\_AutoNodOnSlit"), using jitter between each AB pair to reduce the effect of bad pixels. Each exposure was 150 sec for a cycle total of 10 min. Fainter objects were observed with 2 ABBA cycles.

For each galaxy, an A0 star at a similar airmass was also observed as a telluric standard and flux calibrator with a single AB pair, each frame of 10 sec exposure. Some galaxies shared the same standard star, as they were close together on sky. 

The data were reduced in the standard manner using the \texttt{gasgano}\footnote{\url{http://www.eso.org/sci/software/gasgano.html}} software with the ``sofi\_spc\_jitter'' recipe, with flat-fielding and wavelength calibration. Telluric correction and flux calibration were performed using an A0V stellar spectral template \citep[from the ESO stellar library of][]{Pickles1998}. The spectra were re-dispersed to the range 1495-2518 nm, with 1 nm steps. The velocity resolution is 200 \kms. The spectra were corrected for galactic extinction, using data from the IRSA Galactic Dust Reddening and Extinction service\footnote{\url{https://irsa.ipac.caltech.edu/applications/DUST/}} using the \cite{Schlafly2011} values and the \cite{Cardelli1989} extinction law. Galactic extinction for the catalog objects \ebv{} is in the range 0.007--0.58. They were then wavelength shifted to restframe and cleaned (specifically, blanking the region between the H and K atmospheric windows and also removing single-pixel noise).

One object from the catalogs (WPV85007 = HE0036-5133) had already been observed with SOFI by \cite{Busch2016a}\footnote{\url{http://cdsarc.u-strasbg.fr/viz-bin/qcat?J/A+A/587/A138}} using the same telescope setup; we incorporated the spectrum into our sample. Another object, 6dF-J070841.5-493306 (1H0707-495) had been observed with SINFONI on the VLT, with the data being retrieved from the ESO archive (program ID 092.B-0279(B), PI A.C. Fabian). After standard data reduction using \texttt{gasgano}, we used an aperture of 1.75\arcsec~ on the resulting data cube to extract the galaxy spectrum. This aperture was determined by a curve-of-growth measurement using increasing aperture sizes.

Table \ref{tbl:shirnlsobslog} list a sample of the objects with coordinates, redshift, luminosity distance, galactic extinction and \textit{K} magnitude, plus observation details. The complete table is available online and as supplementary material, as described below.

The pipeline does not give a noise spectrum directly; we can estimate (at least for comparative purposes) a signal-to-noise ratio from the spectrum itself by measuring the average and standard deviation over a spectral window in a ``quiet'' region (i.e. free of atmospheric absorption and spectral features). In general, we use a wavelength of 2000 nm with a window of 100 nm. The resulting values are also given in Table \ref{tbl:shirnlsobslog}, and range from 5--75.

\subsection{Sampling Bias}
\begin{figure*}
\centering
\includegraphics[width=1\linewidth]{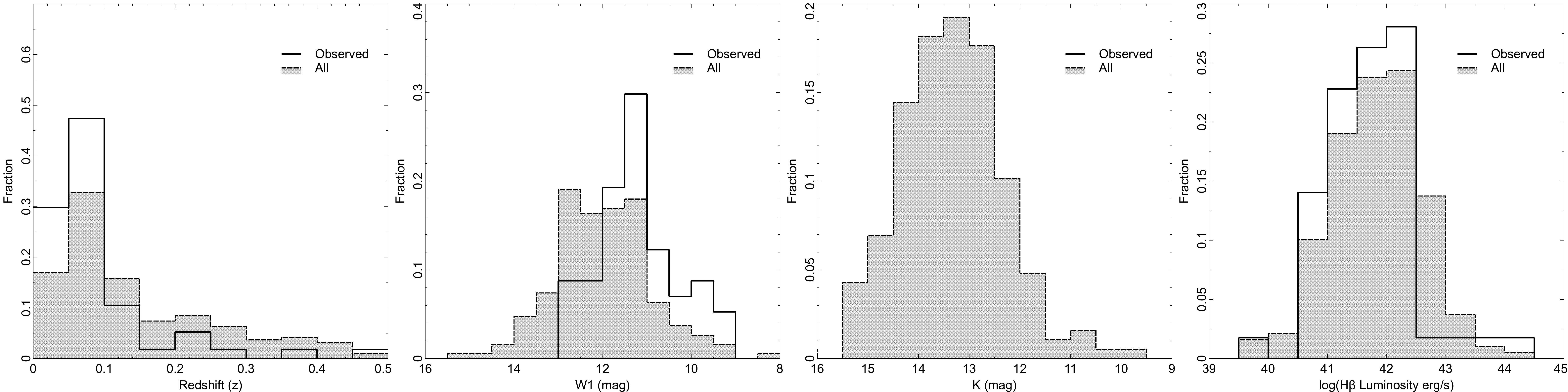}
\caption{Histograms of redshift (\textit{z}), \textit{WISE W1}, 2MASS \textit{K} magnitude and \Hb{} log luminosity of observed objects vs. the full catalog set.} 
\label{fig:shirnlsresults03}
\end{figure*}

We would expect our observations to be a somewhat biased sub-sample of the source catalogs, since the observational priority was given to \textit{K}-band bright objects with strong \Hb{} fluxes. Figure \ref{fig:shirnlsresults03} show the histograms of the observed vs. catalog distributions of redshift, \textit{W1}, \textit{K} magnitude and \Hb{} log luminosity; the observed objects are somewhat brighter in magnitude and closer then the complete sample; the \Hb{} luminosity is somewhat better sampled.
\subsection{Emission Line Fitting - Gaussian vs. Lorentzian Profiles}
Emission lines profiles are typically de-blended into several Gaussian components, which are then equated to individual gas masses with separate kinematics. Both \citetalias{Chen2018} and \citetalias{Schmidt2018} use multiple Gaussian fits to the observed line profiles, as do other surveys \citep[e.g.][the S7 survey]{Dopita2015a}. These fits assume that the narrowest component(s) represents the narrow-line region (NLR) emission, with a single broad component being the BLR emission. However, a Lorentzian fit for the BLR emission (especially for NLSy1's) is more physically motivated, both on observational \citep{Kollatschny2013,Berton2020} and theoretical grounds \citep{Goad2012}, whose models posit a bowl-shaped BLR bounded by a toroidal obscuring region.In their models, the presence of low inclination and significant turbulence leads to narrow cores and extended broad wings in the line profiles.  Similar models have been proposed by \cite{Gaskell2007} and \cite{Gaskell2017}.

As a Lorentzian fit is very similar to a 2-component Gaussian fit, this may lead to a spuriously high value for the NLR emission component. \cite{Berton2020} concludes that the line shape (either Gaussian or Lorentzian) reveals two populations with different black hole masses, Eddington ratios and \OIII{} and Fe II strengths. However, for our purposes, we can use a single Lorentzian fit, since we are comparing two emission lines from each object and the estimates of \mbh{} will not be affected greatly by the specific line shape.

We used the \texttt{QFitsView}\footnote{\url{http://www.mpe.mpg.de/~ott/dpuser/qfitsview.html}} \citep{Ott2012} de-blending functionality to fit all emission lines; this allows for both Gaussian and Lorentzian fits with multiple components. This requires some manual input from the user to set the initial estimates of continuum, height, centre and width and uses the GSL\footnote{\url{https://www.gnu.org/software/gsl/doc/html/index.html}} ``gsl\_multifit'' routines, returning fit values and errors of each component (continuum slope, central wavelength, FWHM and flux).

\begin{figure*}
	\centering
	\includegraphics[width=.8\linewidth]{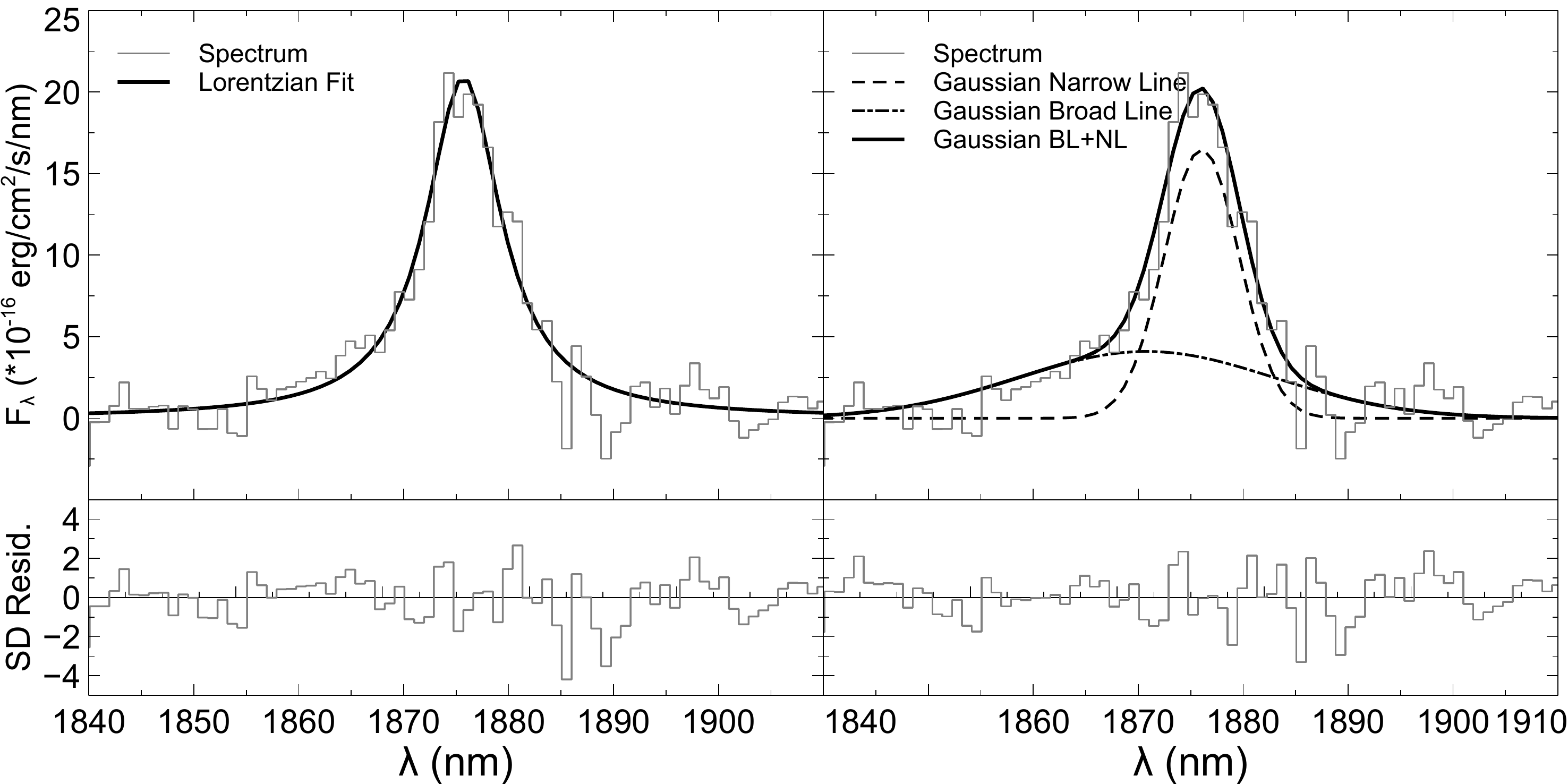}
	\caption{Sample \paa{} emission line of 6dF~J202557.4-482226 fitted with (left panel) a single Lorentzian curve and (right panel) two Gaussian curves. The residuals are plotted at the same vertical scale as the fits.}
	\label{fig:shirnlsresults04}
\end{figure*}

As an example, we fit a single Lorentzian function vs. 2 Gaussians (1 broad + 1 narrow) to the \paa{} spectral line of 6dF~J202557.4-482226 (=~2MASS~J20255738-4822261, \textit{z}=0.06707). These fits are  plotted in Figure \ref{fig:shirnlsresults04}; the residuals are plotted as standard deviations, i.e. the value is divided by the estimated spectral noise (in this case $1.26\times10^{-16}$ \ecs). The residuals from the two fits are of the same size; Occam's razor leads us to prefer the Lorentzian fit (apart from the physical motivation described above). Comparing the fluxes, the Lorentzian fit flux is $2.86\times10^{-14}~$\ecs{}, whereas the combined Gaussian broad and narrow fit flux is $2.73~ (1.26+1.47)\times10^{-14}~$\ecs{}, so \mbh{} will be underestimated if just the broad Gaussian component is used. The dispersion (second moment) of a Lorentzian curve is just the FWHM, whereas for a Gaussian curve $\sigma$=FWHM/2.355.

Several authors \citep{Brotherton1994,Crenshaw2007,Dopita2015a,Schmidt2016,Spence2018a} fit an intermediate velocity component in the BLR emission lines, as well as the standard broad component. Our data show no requirement for this component with Lorentzian fits. On this basis, we will fit the hydrogen broad lines with a single Lorentzian curve.

Of the 57 objects in the NIR sample, we could firmly measure \paa{} in 49 cases. One object (6dF~J195705.3-414117) had \pab{} visible, as the redshift of 0.374 moved \paa{} out of the measurement wavelength range. In this case, the measured \pab{} flux was multiplied by 2.07 (the canonical ratio) to estimate the \paa{} flux.  Two object had poor noise levels and the remaining 5 objects had the \paa{} wavelength in the atmospheric absorption gap between the \textit{H} and \textit{J} bands. We could measure one object with \paa{} in this band, and the line was strong enough to dominate over the noise.

A sample of the flux and FWHM measurement results are given in Table \ref{tbl:shirnlsmeasure} with the derived luminosity, velocity dispersion and black hole mass estimates; these are calculated as discussed in Section \ref{sec:Sec4}. The uncertainties in the flux and FWHM values are 1$sigma$ errors from the fitting routines in \texttt{QFitsView}. The luminosity distance is computed from the redshift using the functionality implemented in TOPCAT\footnote{\url{http://www.star.bristol.ac.uk/~mbt/topcat/}} \citep{Taylor2005}, which depends on the cosmological parameters given above. The velocity dispersions have been corrected for instrumental resolution (6dF (\citetalias{Chen2018}) $R=1000$, REOSC (\citetalias{Schmidt2018}) $R=1580$, SOFI $R=600$) and the flux values are in units of 10\pwr{-15} \ecs.  The complete table is available online and as supplementary material, as described below.

\subsection{Flux Calibration of Optical Spectra}
\label{sec:shirnlsoptfc}
As we wish to measure the hydrogen lines in the optical regime to compare black hole mass calculations and examine extinction in the BLR, we must ensure that the optical flux calibration is accurate. The \citetalias{Schmidt2018} catalog provided flux-calibrated spectra, but \citetalias{Chen2018} used spectra from the 6dFGS survey, which are not flux-calibrated in their raw form. \citetalias{Chen2018} devised a calibration procedure by comparing the 6dFGS spectra with Sloan Digital Sky Survey (SDSS) spectra of the same object; they calculated the ratio of the 6dFGS over SDSS spectra for 7 radio-quiet (therefore minimally variable) cross-matched objects and fitted the averaged spectra with a fifth-order polynomial curve, which was used to calibrate all the spectra. We performed this analysis and obtained the same result. The spectra were then adjusted for galactic extinction as described above, and shifted to restframe. The \citetalias{Schmidt2018} spectra were also extinction-corrected (they had been shifted to restframe before publication).

As a cross-check, we compared the flux-calibrated 6dFGS spectra from \citetalias{Chen2018} with the 6 cross-matched \citetalias{Schmidt2018} spectra. In general, the match for the spectral shape was excellent, however the flux ratios were somewhat scattered over a range of 1/3 to 3, with an average ratio of 1.2. A factor of 2 error in the flux values leads to a difference of 0.25 dex in \mbh{}.

When comparing spectra in different wavelength bands and instruments, the major systematic differences are flux calibrations and aperture effects. The SOFI observations used a 1\arcsec~ slit, whereas the 6dF science fibres have an aperture of 6\arcsec.7. It would be possible to constrain the optical flux calibration by requiring that the optical and NIR spectra be smoothly continuous, but the uncertainties from aperture effects preclude this. We consider that the apertures are large enough to accommodate the (presumably) point-source BLR, and thus to capture the whole emission-line radiation.

\section{Results}
\label{sec:Sec4}
\subsection{Black Hole Mass Estimation}
\subsubsection{Mass Estimate by \LBCont{} vs. \Hb{} Luminosity}
In general, \mbh{} is estimated from a relationship which assumes virialised gas clouds in the BLR:
\begin{equation}
    \mmbh = f\frac{v^2 R_{BLR}}{G}
\end{equation}
where $R_{BLR}$ is the radius of the BLR, $v$ is the gas velocity, $G$ is the gravitational constant and $f$ is a factor dependant on the BLR geometry. The FWHM of the \Hb{} line is a proxy for the rotational velocity of the gas, and the luminosity of the continuum at 5100 \AA{} (\LBCont) is a proxy for the $R_{BLR}$, proportional to $R^2$.  \cite{Greene2005} also showed that \Ha{} (and thus \Hb) broad line emission has a tight relationship with \LBCont.  

\citetalias{Chen2018} based their estimation of \mbh{} of their catalog objects on the \LBCont{} continuum luminosity and the dispersion of the \Hb{} broad line \citep{Foschini2015,Bentz2013,Berton2015}. \citetalias{Schmidt2018} used the methods of \cite{Greene2005} (i.e. based mainly on the \Ha{} line attributes, their Eqns. 6 and 7); of the 28 galaxies in the catalog, 26 use the \Ha{} line and 2 use the \Hb{} line. \cite{Kim2010} derived black hole mass estimates from \paa{} lines, comparing with those derived from either reverberation mapping method or single-epoch measurement method using Balmer lines. For consistency, we will use the relationships from \cite{Kim2010}. 

The relationship for \mbh{} with emission line parameters is:
\begin{equation}
	\label{eqn:shirnlsmbhcalc}
	\mmbh (M_\odot) = \alpha\left(\frac{L_{H}}{10^{42}~erg~s^{-1}}\right)^\beta\left(\frac{FWHM_{H}}{1000~km~s^{-1}}\right)^\gamma
\end{equation}
where $L_H$ and $FWHM_H$ are the luminosity and FWHM of species $H$ (\paa{} or \Hb), and the constants are given in Table \ref{tbl:shirnlsmbhconst}; we follow the recommendation of \cite{Kim2010} to use the relationship which has the more physical exponent of the FWHM fixed at $\gamma=2$. We note that the value of $\beta$ is very close to the theoretical value of 0.5.
 
\begin{table}
\centering
	\caption{BH Mass Calculation Constants}
	\label{tbl:shirnlsmbhconst}
	\begin{tabular}{@{}lccc@{}}
	\toprule
& $\alpha$             & $\beta$             & $\gamma$ \\ \midrule
	\paa & $7.16\pm0.4$  & $0.49\pm0.06$ & 2 \\
	\Hb  & $6.88\pm0.57$ & $0.46\pm0.05$ & 2 \\ \bottomrule
	\end{tabular}
\end{table}

\begin{figure*}
	\centering
	\includegraphics[width=.8\linewidth]{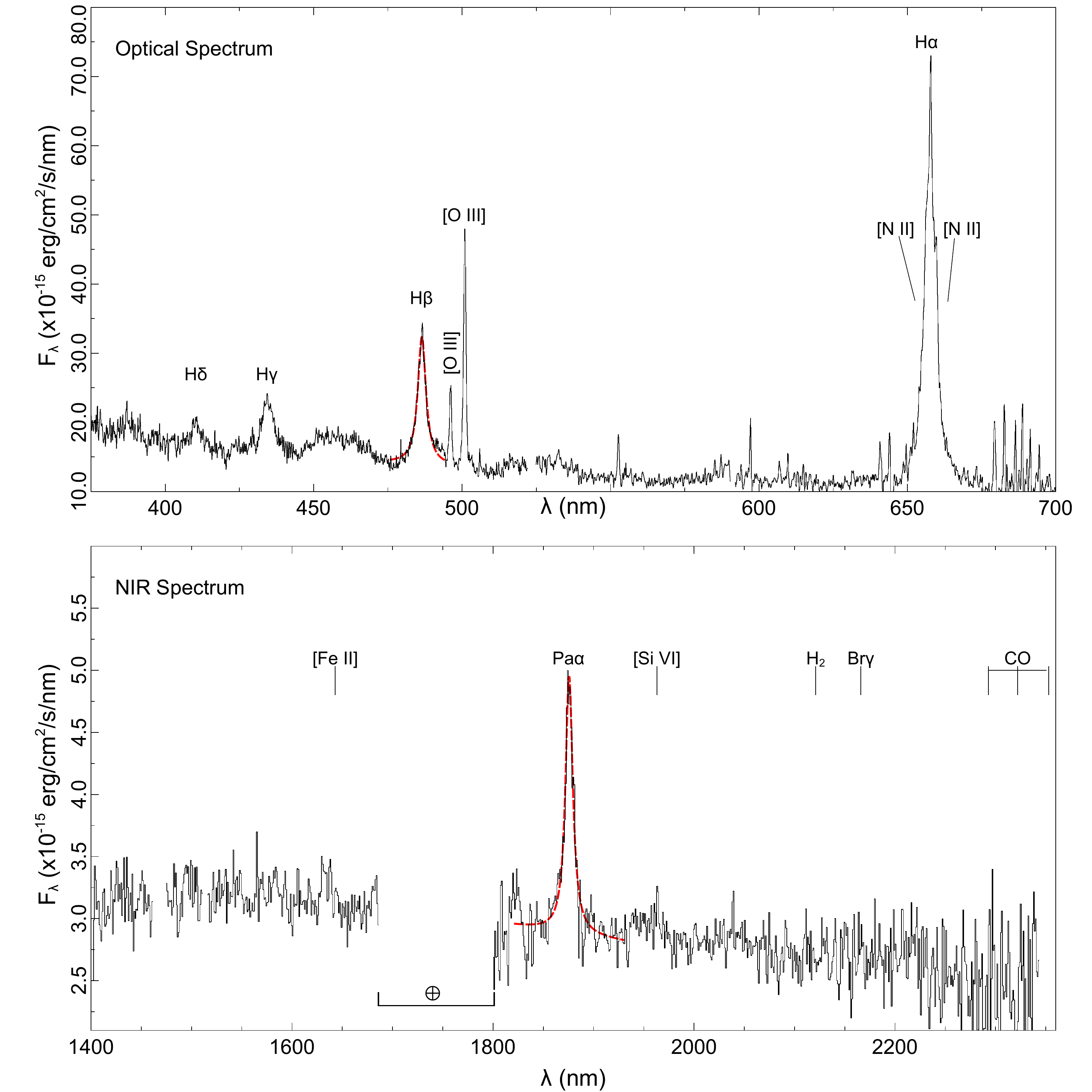}
	\caption{Optical and infrared restframe spectra of galaxy 6dF J2025574-482226. Emission lines are labelled. The \textbf{$\oplus$} symbol on the NIR spectrum is the telluric absorption band between the \textit{H} and \textit{K} atmospheric windows. The Lorentzian fits to the \Hb{} and \paa{} emission lines are shown in red.}
	\label{fig:shirnlsresults05}
\end{figure*}

\begin{figure*}
	\centering
	\includegraphics[width=.7\linewidth]{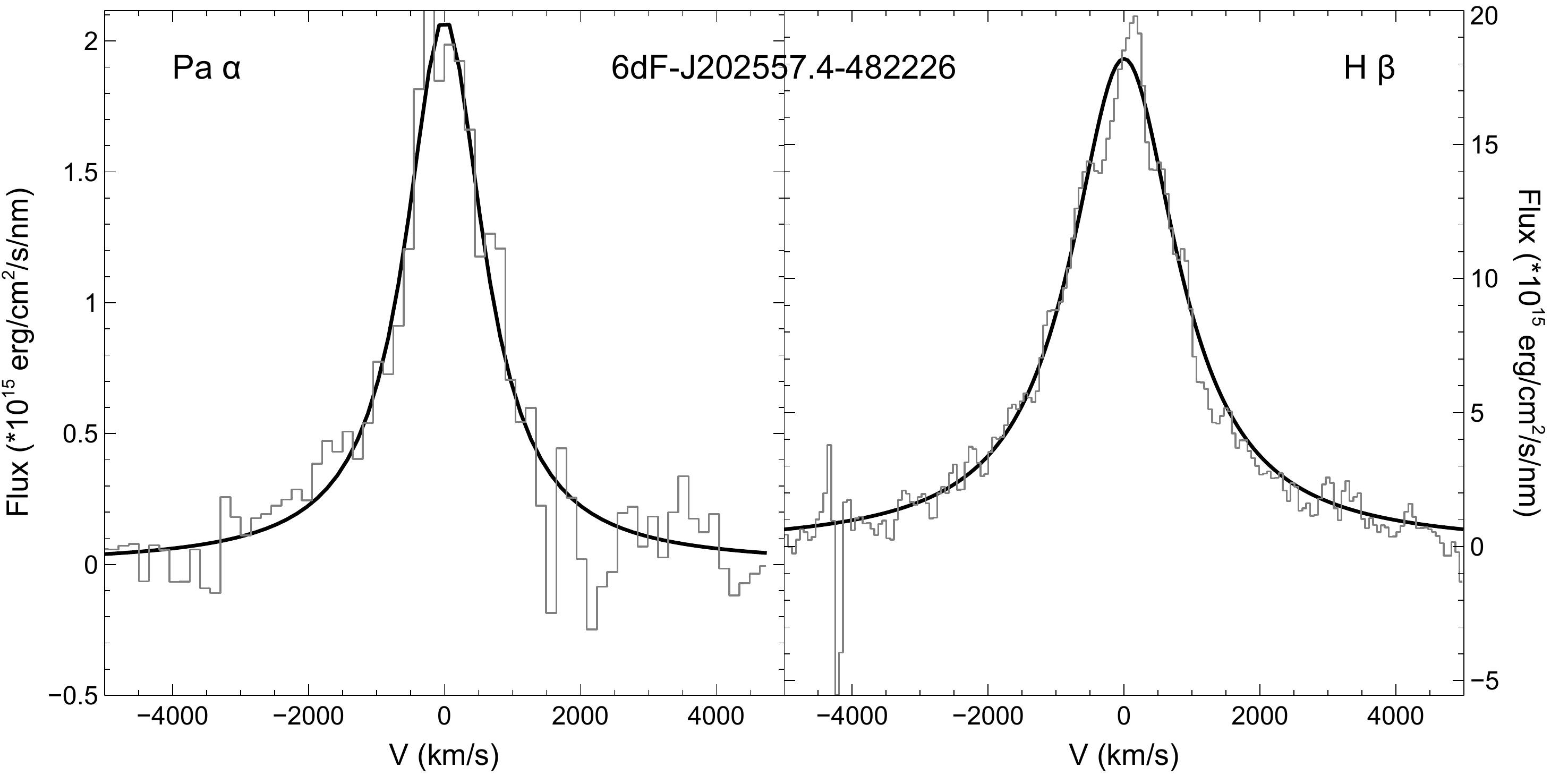}
	\caption{Line-profile fitting (using a Lorentzian curve) of \Hb{} and \paa{} emission lines from galaxy 6dF J2025574-482226. The fluxes are in units of 10$^{-16}~$ \ecs nm\pwr{-1}, and the spectra are continuum subtracted.}
	\label{fig:shirnlsresults06}
\end{figure*}

As an example, we show a plot of the optical and infrared spectra of the object 6dF~J2025574-482226, with the Lorentz fits on both the \Hb{} and \paa{} lines (Figure \ref{fig:shirnlsresults05}). Figure \ref{fig:shirnlsresults06} shows the fits compared in greater detail where the continuum is removed and the wavelength is converted to relative velocity. As can be seen, the \paa{} spectral line is somewhat narrower than the \Hb{} line; this will be discussed below. A complete set of plots of the line fits for all the 48 objects with both \paa{} and \Hb{} flux is available online as as supplementary material.
\subsubsection{Recalculated Mass from Optical Measurements}
\begin{figure*}
	\centering
	\includegraphics[width=1\linewidth]{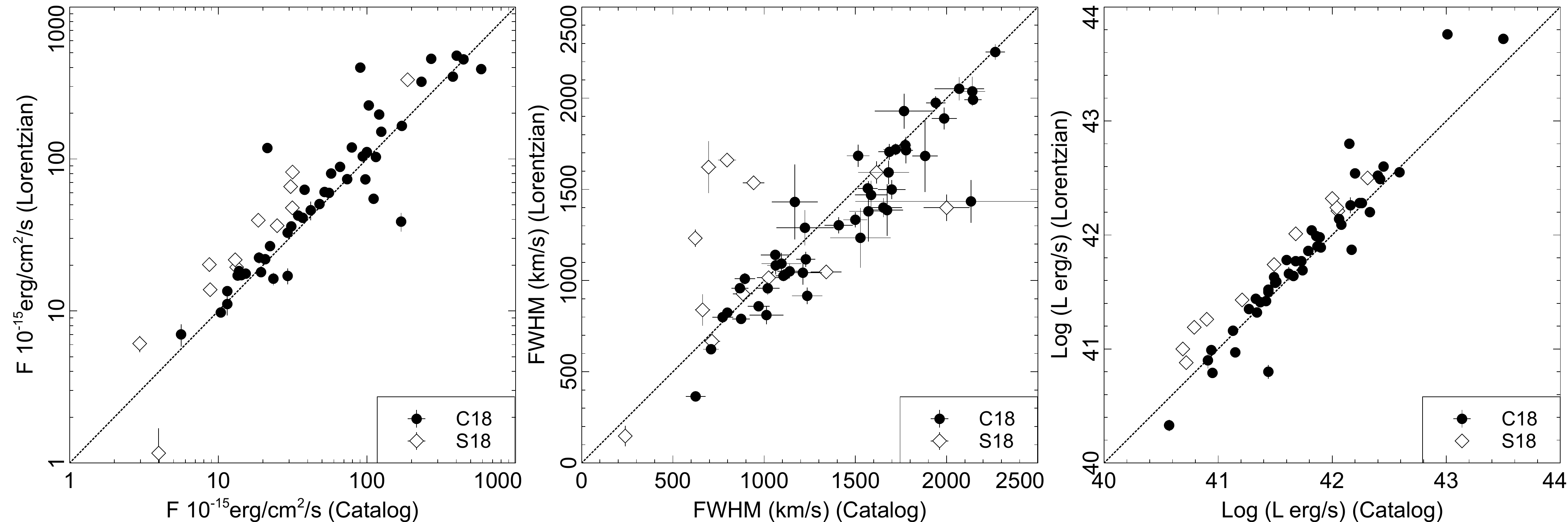}
	\caption{Comparison of \citetalias{Chen2018} and \citetalias{Schmidt2018} catalog data  with remeasured Lorentzian fit for \Hb{} flux(left panel), FWHM velocity dispersion (middle panel) and log luminosity (right panel).}
	\label{fig:shirnlsresults07}
\end{figure*}

We re-measured the \Hb{} broad line with a single Lorentzian curve for all objects that had measurable \paa{} flux in our survey (see Section \ref{sec:observations}) using the \texttt{QFitsView} functionality, and recomputed \mbh{} with the measured flux and FWHM. The line widths were corrected for instrumental dispersion. We compared the catalog data with the flux, luminosity and FWHM values derived from the Lorentzian fit; Figure \ref{fig:shirnlsresults07} shows this.  On these plots, the diagonal lines denote the equality between the measures; while the broad component fluxes and luminosities are very similar, the FWHM measurements are somewhat more scattered. This is more prevalent in the \citetalias{Schmidt2018} data; this is probably due to (1) measuring \Hb{} instead of \Ha, and (2) the difficulty of de-composing the \Ha{} from the associated \NII{} lines. 
\begin{figure*}
	\centering
	\includegraphics[width=1\linewidth]{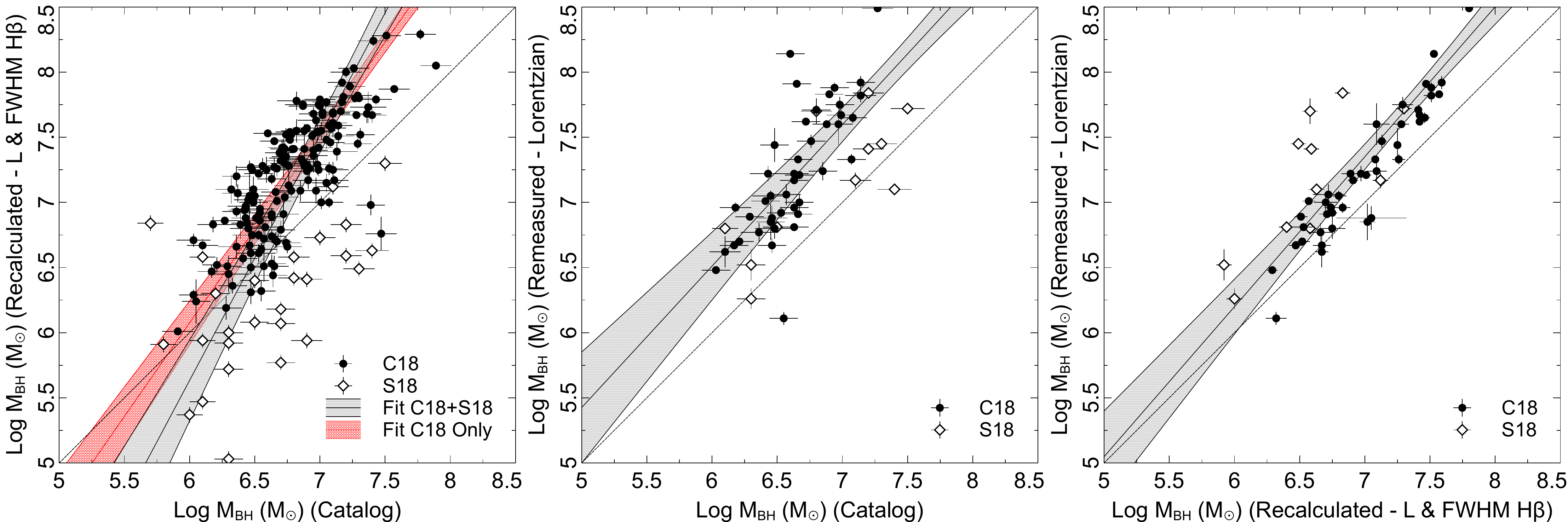}
	\caption{(Left panel) \mbh{} values from optical data of the original catalog value vs. the recalculated value, for all objects in the original catalogs. (Middle panel) Catalog vs. remeasured Lorentzian fit \mbh{}. (Right panel) Remeasured vs. Lorentzian fit \mbh{}. The orthogonal least-squares fit lines with the 1$\sigma$ confidence intervals are shown; for the left panel, these are the combined (black) and \citetalias{Chen2018} alone (red) data.}
	\label{fig:shirnlsresults08}
\end{figure*}
Using the catalog values of \Hb{} flux and FWHM, we can re-derive the \mbh{} using the method given in \cite{Kim2010} and compare that with the \mbh{} of the original catalogs (Figure \ref{fig:shirnlsresults08} - left panel).

We performed an orthogonal least squares regression analysis using the BCES method of \cite{Akritas1996}\footnote{The BCES routine was from the Python module written by Rodrigo Nemmen \citep{Nemmen2012}, which is available at \url{https://github.com/rsnemmen/BCES}}; this regression method is preferred because there is no independent variable and both data sets have errors. The best-fit slope for the catalog vs. recalculated \mbh{} (Figure \ref{fig:shirnlsresults08} - left panel in black) is:
\begin{equation}
\begin{split}
    \log{\mmbh}(\rm{Recalc}) = 1.88\pm0.15 \times log{\mmbh}(\rm{Cat})\\
    -5.65\pm1.02
\end{split}
\end{equation}
We note that the \citetalias{Schmidt2018} masses are more scattered and somewhat below the fit line; this is probably due to the difficulties of measuring the \Ha{} FWHM, as discussed before. If we fit just the \citetalias{Chen2018} data, the fit is (Figure \ref{fig:shirnlsresults08} - left panel in red):
\begin{equation}
\begin{split}
    \log{\mmbh}(\rm{Recalc}) = 1.44\pm0.08 \times log{\mmbh}(\rm{Cat})\\
    -2.55\pm0.56 
\end{split}
\end{equation}
This is somewhat closer to a slope of unity; see the next section for discussion of this.

For the catalog vs. Lorentzian remeasured \mbh{} (Figure \ref{fig:shirnlsresults08} - middle panel), the fit is:
\begin{equation}
\begin{split}
    \log{\mmbh}(\rm{Remeasured}) = 1.09\pm0.01 \times log{\mmbh}(\rm{Cat})\\ -4.448\pm1.553 
\end{split}
\end{equation}
For the recalculated vs. Lorentzian remeasured \mbh{} (Figure \ref{fig:shirnlsresults08} - right panel), the fit is:
\begin{equation}
\begin{split}
    \log{\mmbh}(\rm{Remeasured}) = 1.15\pm0.07 \times log{\mmbh}(\rm{Recalc})\\ -0.004\pm0.002 
\end{split}
\end{equation}
\subsubsection{Malmquist Biases}
\begin{figure*}
	\centering
	\includegraphics[width=1\linewidth]{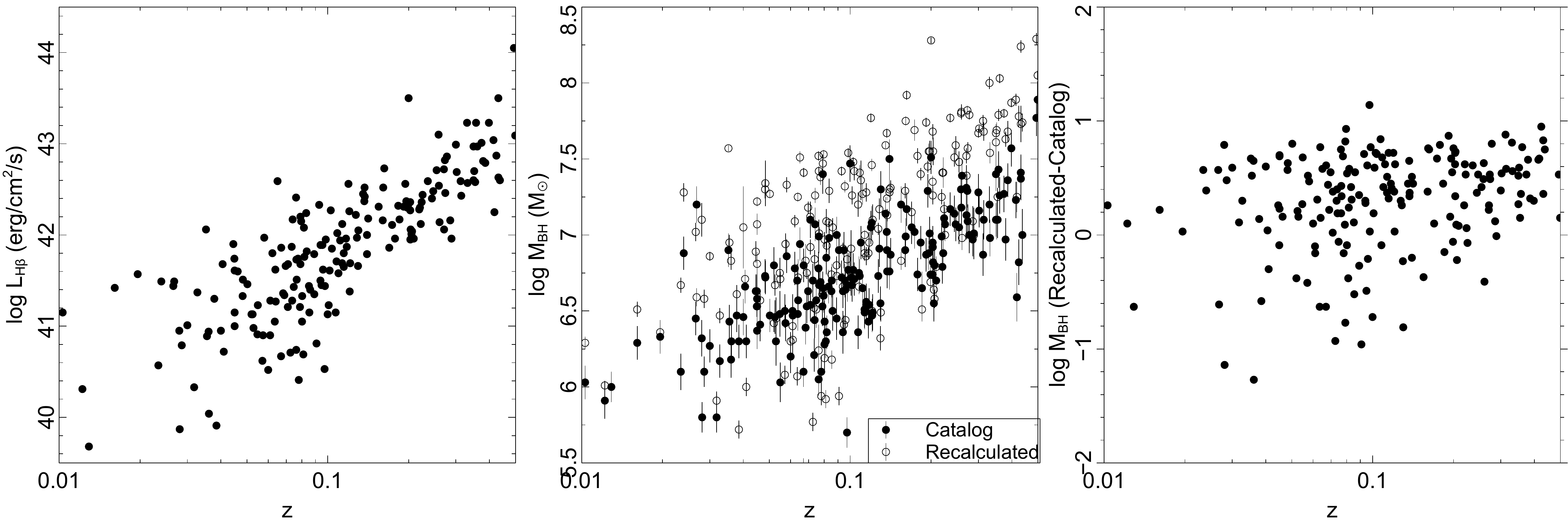}
	\caption{Catalog luminosity and \mbh{} vs. redshift, showing the Malmquist bias of the flux-limited sample, plus the trend of the mass ratios with redshift caused by stellar light contamination. (Left panel) Log luminosity (Middle panel) \mbh, with both catalog and recalculated values (Right panel) Ratio of catalog to recalculated \mbh.}
	\label{fig:shirnlsresults09}
\end{figure*}

As noted above, the masses recalculated by the \cite{Kim2010} method  (Figure \ref{fig:shirnlsresults08} - left panel) are somewhat higher than the catalog values, with the slope of the fit greater than unity. A reason for this is the contamination of the \LBCont{} by stellar light at lower redshifts. In a flux-limited survey, Malmquist biases can play an important role; as objects are further away, only the brighter objects are detected, plus with the greater volume, more extreme objects with lower space density are included. This is illustrated in Figure \ref{fig:shirnlsresults09} (left panel) where both the lower and upper limits of the \Hb{} luminosity increase with redshift. In a similar manner, the derived \mbh{} (middle panel) also shows this trend; we also show the differences between the values derived from the \LBCont{} method, vs. the ``pure'' \Hb{} method. The right panel shows the difference between the two methods; there is a trend to larger differences with redshift, as the objects selected have more luminous and massive AGNs, which have a greater contribution to \LBCont{} over the stellar light. 
\subsection{\Hb{} and \paa{} Line Profiles Comparison}
In comparing the parameters of the \Hb{} and \paa{} emission lines, we note that these both originate from the same upper level (transitions 4--2 and 4--3), which means they reflect the same excitation conditions. The flux ratios \Hb/\paa for Case B depend on the temperature and density of the emitting gas; the standard values used for most AGN studies are $T_e = 10^4$K and $n_e = 10^{3}$ cm\pwr{-3}, which gives a ratio of $\sim3$ \citep{Hummer1987}. This ratio is a weak function of both temperature and density, and will rise to a value of $\sim3.72$ at $T_e = 3\times10^4$K and $n_e = 10^{9}$ cm\pwr{-3}, more typical of BLR conditions. AT higher densities, the ratio asymptotes to 3.62, with no temperature dependence. Significantly lower measured flux ratios are, in the first instance, caused by dust extinction, which can be structurally complex in the extreme environments of AGN. The dust may be distributed throughout the central engine, or may be from galactic obscuration. The line width, however, will reflect the velocity structure of the emitting regions. 

\begin{figure*}
	\centering
	\includegraphics[width=1\linewidth]{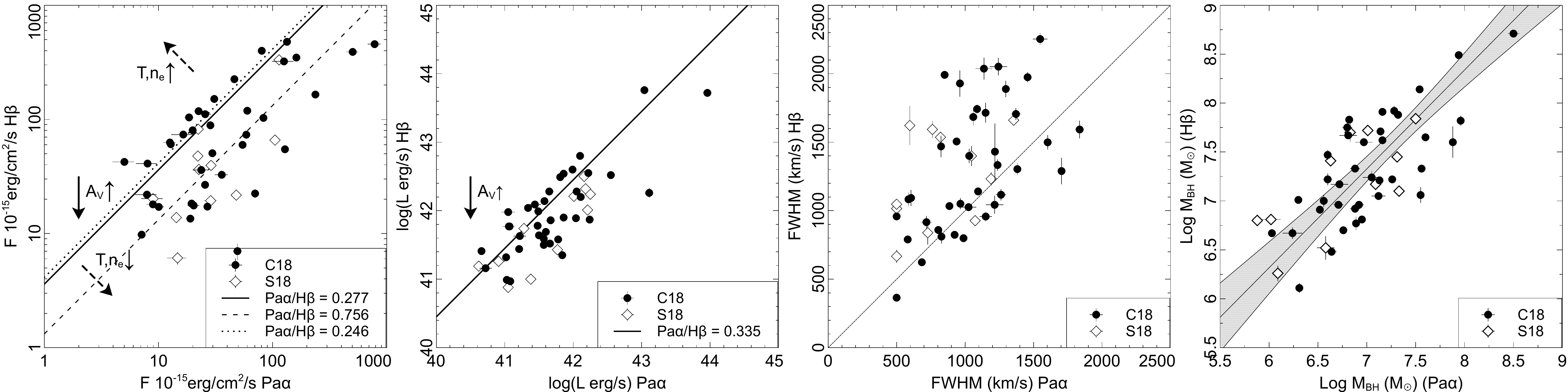}
	\caption{Comparison of \Hb{} and \paa{} measurements. Panel 1:  emission line flux, panel 2: luminosity, panel 3: FWHM velocity, panel 4: \mbh.}
	\label{fig:shirnlsresults10}
\end{figure*}

\begin{figure*}
	\centering
	\includegraphics[width=1\linewidth]{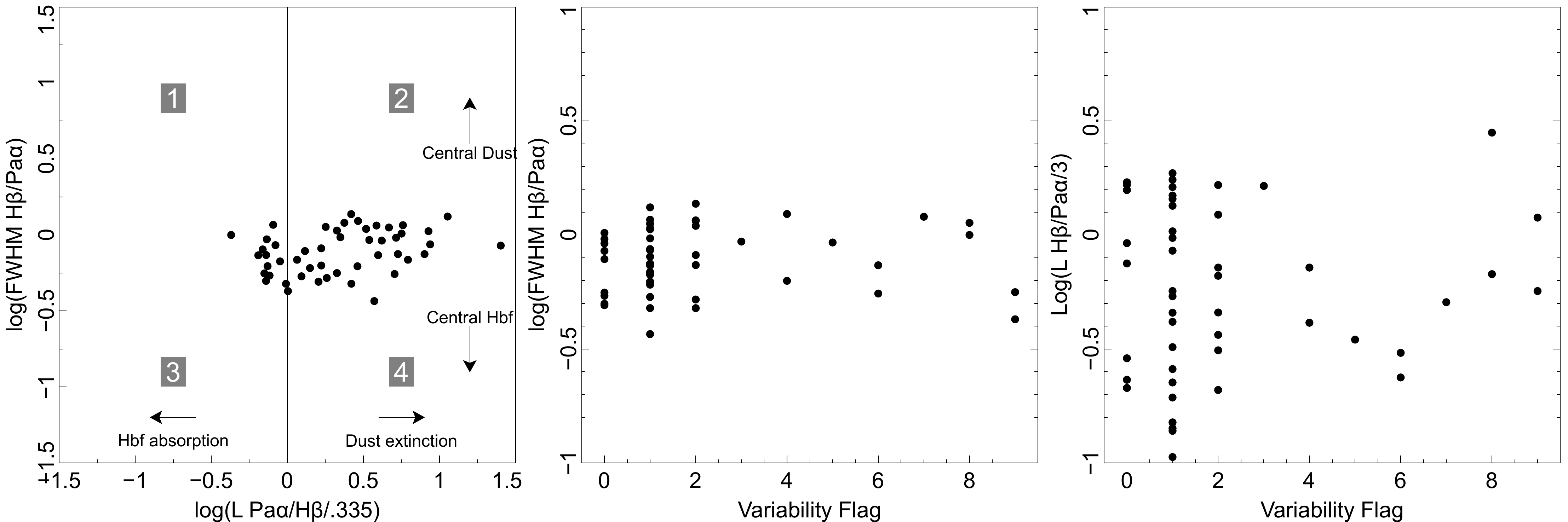}
	\caption{Comparison of \Hb{} and \paa{} luminosity and FWHM ratios (left panel) and with \textit{WISE} variability flag (middle and right panels). The FWHM and luminosity ratios are plotted on a log scale, and the luminosity ratios are divided by the expected flux ratio of 0.277. On the luminosity vs. FWHM plot, the regions are numbered for further discussion. }
	\label{fig:shirnlsresults11}
\end{figure*}

We compare the line flux, luminosity and FWHM velocity as well as \mbh{} from the \paa{} and \Hb{} emission lines: this is shown in Figure \ref{fig:shirnlsresults10}. The luminosities are reasonably correlated, clustering around the fiducial line with the ratio \paa/\Hb{}~=~0.335. For reference, the lines indicating the ratios expected for the range of physical condition are plotted, from 0.756 (low density and temperature) to 0.246 (high density and temperature). The diagonal dashed arrows 
represent the directions of the lines for lower or higher density and temperature. The values of \Hb{} luminosity that lie below  the fiducial lines can be attributed to dust extinction (increasing $A_V$, as indicated by the arrow). The extinction will be discussed further in section \ref{sec:shirnlsextblr}.
The orthogonal least-squares fit on the computed BH masses (Figure \ref{fig:shirnlsresults10} - panel 4) is:
\begin{equation}
\begin{split}
    \log{\mmbh}(\rm{H\beta}) = (1.011\pm0.119) \times log{\mmbh}(\rm{Pa\alpha})\\ -(0.248\pm0.831)    
\end{split}
\end{equation}
This is compatible (within measurement errors) of equivalence of the calculated mass by the two methods.
We check whether the differences between the emission-line luminosity and FWHM are correlated by plotting the respective ratios, as shown in Figure \ref{fig:shirnlsresults11} (left panel). The ratios are plotted in log scale, with the flux ratio adjusted to the expected fiducial ratio (with the lines indicating equality). No correlation can be observed; from which we deduce that the differences are not due to a measurement bias. The quadrants are numbered for further discussion below, with arrows representing the absorption modes and locations.

We also check the source variability; we use the \textit{WISE} catalog variability flag, ranging from 0 to 9. We plot both the luminosity and FWHM ratios against this flag (Figure \ref{fig:shirnlsresults11} - middle and right panels) and again find no relationship. This implies that source variability between the \Hb{} and \paa{} measurement epochs is not the reason for any scatter in the ratios.
\subsection{Extinction in the BLR}
\label{sec:shirnlsextblr}
Knowledge of the reddening of AGN is crucial for studies of their nature, demographics and evolution \citep{Gaskell2017}.
We can investigate the extinction by dust in the BLR of our objects by examining the ratios of the broad permitted hydrogen lines (\Ha, \Hb{} and \paa) in both the optical and infrared spectra. Similar studies have been done by \cite{Kim2010} and \cite{Schnorr-Muller2016a} (using \Ha, \Hb, \Hg, \pab, \pag{} and \pad). The observed spectra have been corrected for Galactic extinction (see Section \ref{sec:observations}), so any further extinction is intrinsic to the object. In general, the extinction \ebvsub is calculated from the ratio of the flux of two hydrogen species using the formula:
\begin{equation}
\label{eqn:shirnlsextnc}
    E_{B-V}~=~\alpha~\log\left(\frac{f_{1}/f_{2}}{\beta}\right)
\end{equation}
where $f_{1}, f_{2}$ are the fluxes of species 1 and 2, $\alpha$ is a constant depending on the extinction law and the wavelengths of the two species and $\beta$ is the intrinsic ratio of the fluxes given the physical conditions of the emitting region. 

The literature values of $\beta$ for the Balmer decrement (\Ha/\Hb) are disparate; \cite{Schnorr-Muller2016a}, from their extinction study using multiple hydrogen lines, has a wide range (2.5--6.6), \cite{Kim2010} gives 3.33 (derived from their ratios \Ha/\paa~=~9.00 and \Hb/\paa~=~2.70) and \cite{Gaskell2017} concludes, from a study of blue AGNs (which are presumably unreddened), that the hydrogen line ratios follow Case B and that \Ha/\Hb{}~=~2.72. We will use the \cite{Hummer1987} values for high density ($N_e = 10^9$ cm\pwr{-1}); these are relatively insensitive to temperature, but we will use $T_e = 10^4~K$. We will also use the \cite{Cardelli1989} extinction law, with the modifications in the functional shape in the optical from \cite{ODonnell1994}; Table \ref{tbl:shirnlsextnc} gives the values for $\alpha$ and $\beta$ for the emission lines. (For convenience, we use the longer wavelength line as the numerator in the flux ratio, as an increase in this ratio implies an increase in extinction).
\begin{table}
\centering
\caption{Parameter values for Equation \ref{eqn:shirnlsextnc}. The emission line wavelengths are in nm, and the parameter values ($\alpha$ and $\beta$) are discussed in the text}
\label{tbl:shirnlsextnc}
\begin{tabular}{@{}llrrrr@{}}
\toprule
Species 1 & Species 2 & $\lambda_1$ & $\lambda_2$ & $\alpha$ & $\beta$ \\ \midrule
\Ha       & \Hb       & 656.3       & 486.1       & 2.18     & 2.62    \\
\paa      & \Hb       & 1875.6      & 486.1       & 0.78     & 0.269   \\ \bottomrule
\end{tabular}
\end{table}

\begin{figure*}
	\centering
	\includegraphics[width=.8\linewidth]{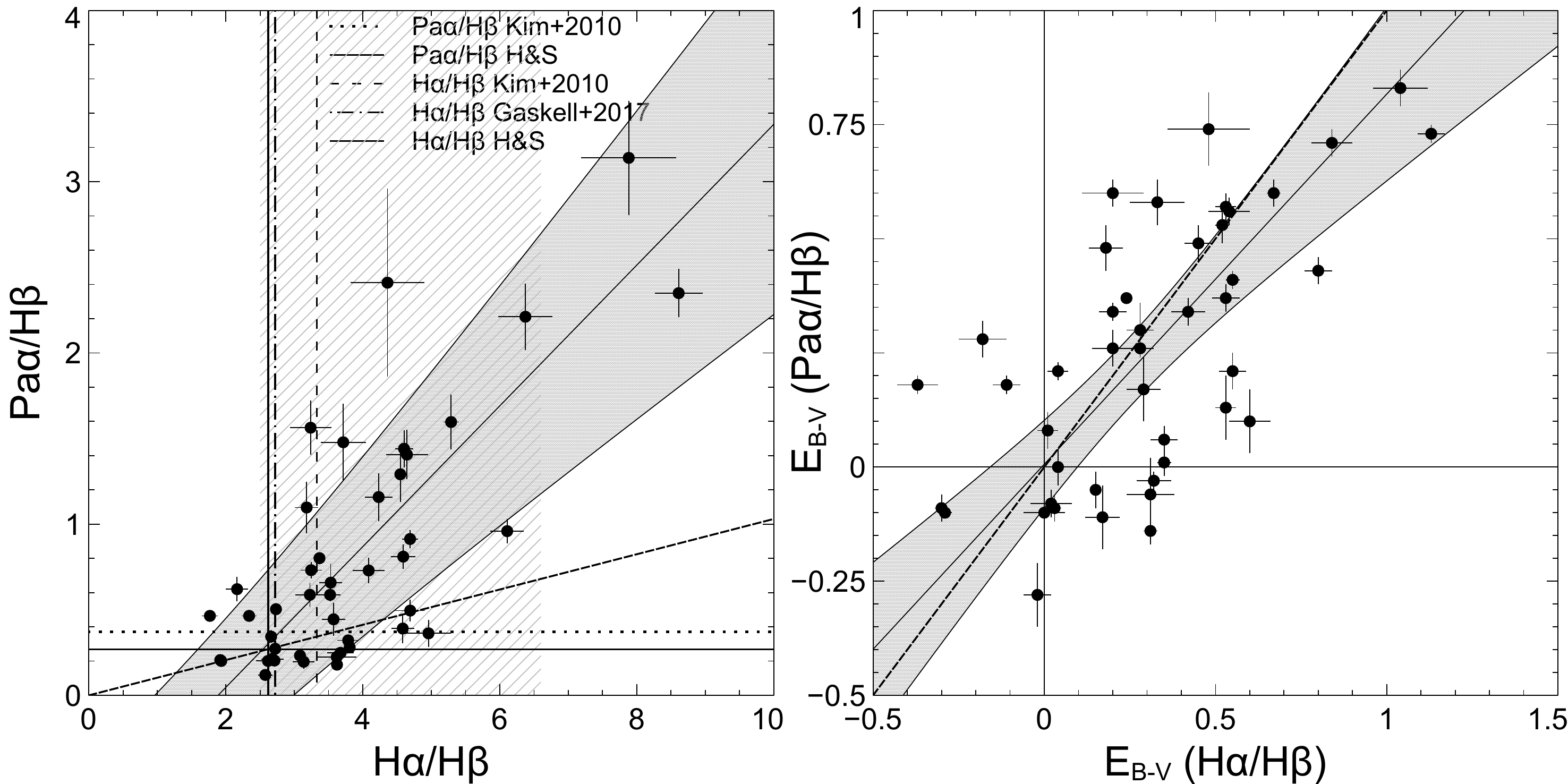}
	\caption{Flux ratios (left panel) and relative extinctions (right panel) for the two extinction measure derivations. For the flux ratios, the fiducial lines show the expected ratios for zero extinction as given in the legend.  \citep[The shaded area shows the range from][for \Ha/\Hb{} of 2.5--6.6]{Schnorr-Muller2016a}. The orthogonal least-squares fits to the data are shown with confidence intervals. The dashed line plots equal extinction from each method, i.e. (i.e. $E_{B-V}^{Pa\alpha}~=~E_{B-V}^{H\alpha}$.}
	\label{fig:shirnlsresults14}
\end{figure*}

Of the 57 observed objects, we could measure \Ha{} from the optical spectra in 51 cases. The other 6 had the \Ha{} line red-shifted beyond the spectral range. Of these, the \paa{} line was not measurable in 7 spectra, leaving 44 objects with \Ha, \Hb{} and \paa{}. The \Ha/\Hb{} and \paa/\Hb{} ratios are plotted against each other in Figure \ref{fig:shirnlsresults14} (left frame). We also plot the fiducial lines for no extinction for each of the references mentioned above. The derived relative extinctions \ebvsub{} from \Ha/\Hb{} ($E_{B-V}^{H\alpha}$) and \paa/\Hb{} ($E_{B-V}^{Pa\alpha}$) are plotted in Figure \ref{fig:shirnlsresults14} (right frame). For both frames, we plot the orthogonal least-squares fits with confidence intervals, as well as the locus of equal extinctions from each method (i.e. $E_{B-V}^{Pa\alpha}~=~E_{B-V}^{H\alpha}$). The fits are:
\begin{align}
    \rm{Pa}\alpha/\rm{H}\beta&=~(0.411\pm0.059)\times\rm{H}\alpha/\rm{H}\beta + (0.775\pm.208)\\
    E_{B-V}^{Pa\alpha}&=~(0.811\pm0.123)\times E_{B-V}^{H\alpha}+(0.006\pm0.056)
\end{align}
We note that the fit to the extinctions gives a slope of unity from each method, within errors. There is a fair degree of scatter about the fit lines (up to 0.5 mag in \ebvsub), and some points show ratios below those expected, producing un-physical `negative' extinction measures for one or both methods, with minimum ratios of 0.12 (\paa/Hb) and 1.77 (\Ha/\Hb). For the \paa/Hb{} ratio, this could be caused by flux calibration differences between the optical and infrared spectra; for the \Ha/\Hb{} ratio, this could be caused by the 6dF calibration method described in Section \ref{sec:shirnlsoptfc} or other measurement uncertainties. Alternately, the emission could be generated in physical conditions that are far from those described above. In Section \ref{sec:Sec5}, we present mechanisms for differential absorption. Note that we have not corrected the estimated BH masses for extinction, given the disparity of values between the two calculation methods.

\subsection{Limits on NLR Emission}
In the near infrared, the species \Fe{} and \Htwo{} can be used as diagnostics of excitation mechanisms, distinguishing between photo-ionization and shock excitation. Following the limits defined in \cite{Riffel2013a}, excitation diagram regimes for AGNs have flux ratios of \Htwo/\brg $>$ 0.4 and \Fe($\lambda1257$)/\pab{} $>$ 0.6. Thus, the flux values of \Htwo{} and \Fe{} put upper limits on the hydrogen emission fluxes in the NLR. Using the standard flux ratios of \Fe$\lambda1257/\Fe\lambda1644 = 1.36$ \citep{Nussbaumer1988}, \paa/\pab~=~1.82 and \paa/\brg~=~10.6 \citep{Hummer1987}, we derive a limit of \Fe($\lambda1644$)/\paa{}~$>$~0.242 and \Htwo/\paa~$>$~0.038, in the absence of significant extinction in the IR.

\begin{figure}
	\centering
	\includegraphics[width=1\linewidth]{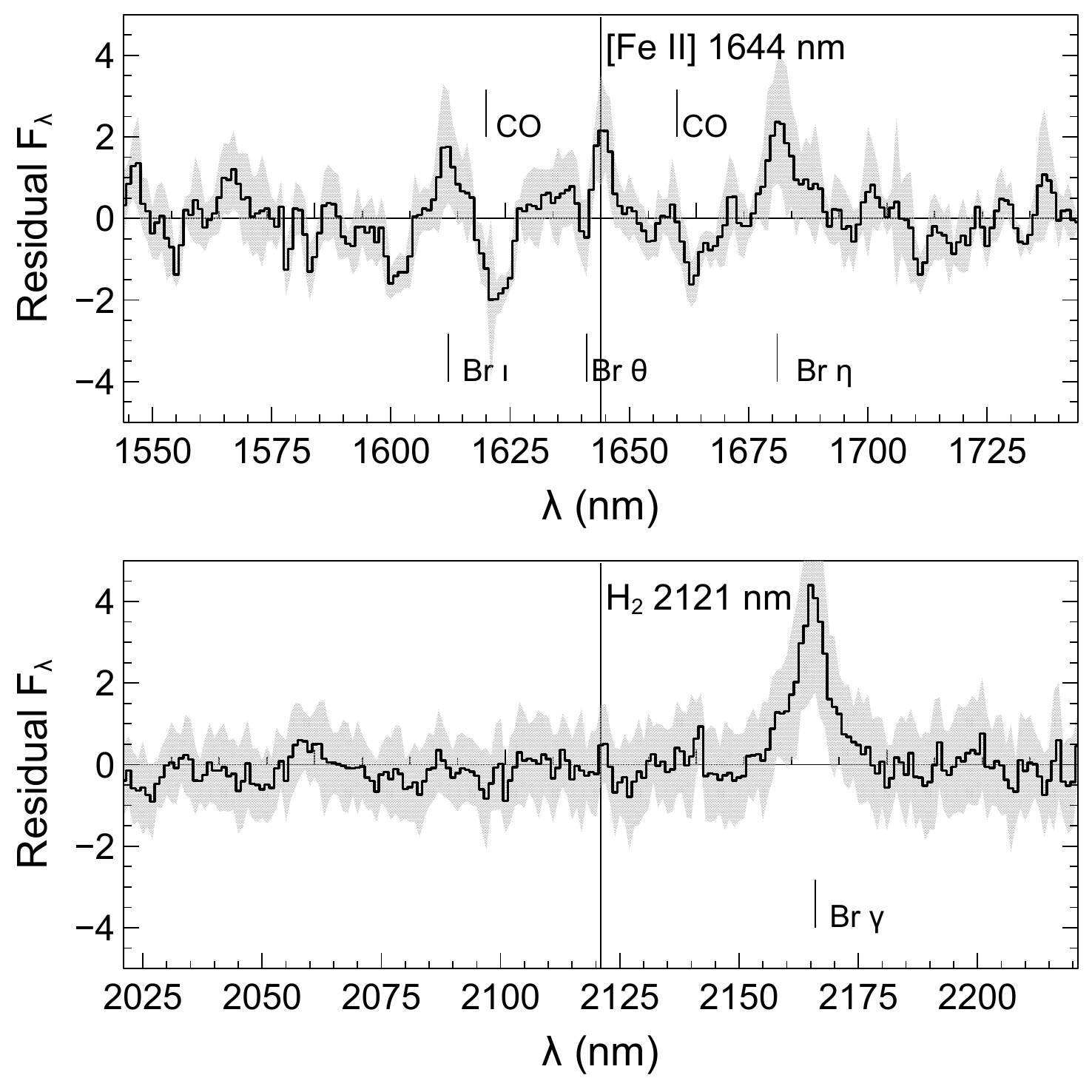}
	\caption{Average stacked spectra of the \Fe{} 1644 nm and \Htwo{} 2121 nm emission lines. The spectra have been continuum subtracted. The Y axes are in units of 10\pwr{-16} \ecs nm\pwr{-1}. No \Htwo{} emission is visible, with only minimal \Fe (which is probably blended with Br $\theta$). Other emission and absorption lines have been marked.}
	\label{fig:shirnlsresults16}
\end{figure}

None of the individual spectra had measurable \Fe{} or \Htwo{} flux above the noise level. To confirm this, we stacked the spectra over a 200 nm window around the line wavelength. For each spectrum we subtracted a continuum fitted with a 5th order polynomial; the spectra are then averaged. This is plotted in Figure \ref{fig:shirnlsresults16}; an emission line is visible (with a flux of $\sim 9 \times 10^{-16}$ \ecs), but is in a region of multiple absorption (e.g. CO at 1620, 1640 and 1660 nm) and emission (Br $\iota$ 1611.4 nm, Br $\theta$ 1641 nm and Br $\eta$ 1681 nm), increasing the uncertainty of flux - the \Fe{} line can blend with Br $\theta$. The \Htwo{} line is not visible, with the \brg{} line at 2166 nm prominent. The errors for the stacked spectral points are estimated from the median deviation of the flux density values at each wavelength.

We consider that the NLR emission of \Fe{} and \Htwo{} (and thus of narrow-line hydrogen emission) is not visible because, given the nearly pole-on view of the BLR, the AGN light overwhelms the NLR light. This is the point of difference with those studies that fit both broad and narrow line emission components to the permitted BLR lines, and supports the single Lorentzian fit paradigm.

\section{Discussion}
\label{sec:Sec5}
\subsection{Differential Absorption by Clouds}
The line structure is considered to be formed by emission line clouds, moving turbulently in a gravitational potential, whose circular velocity is close to Keplerian and dominated by the SMBH. Line profiles formed at an optical depth closer to the SMBH will have a higher velocity dispersion and line width. Inspection of Figure \ref{fig:shirnlsresults10} (panel 3) shows that FWHM of \paa{} is mostly below that of \Hb{}; this accords with the findings of \cite{Kim2010}, suggesting that the Paschen and Balmer broad lines originate from a similar BLR, but with Balmer lines coming more from the inner region of the BLR. 

\begin{figure*}
	\centering
	\includegraphics[width=1\linewidth]{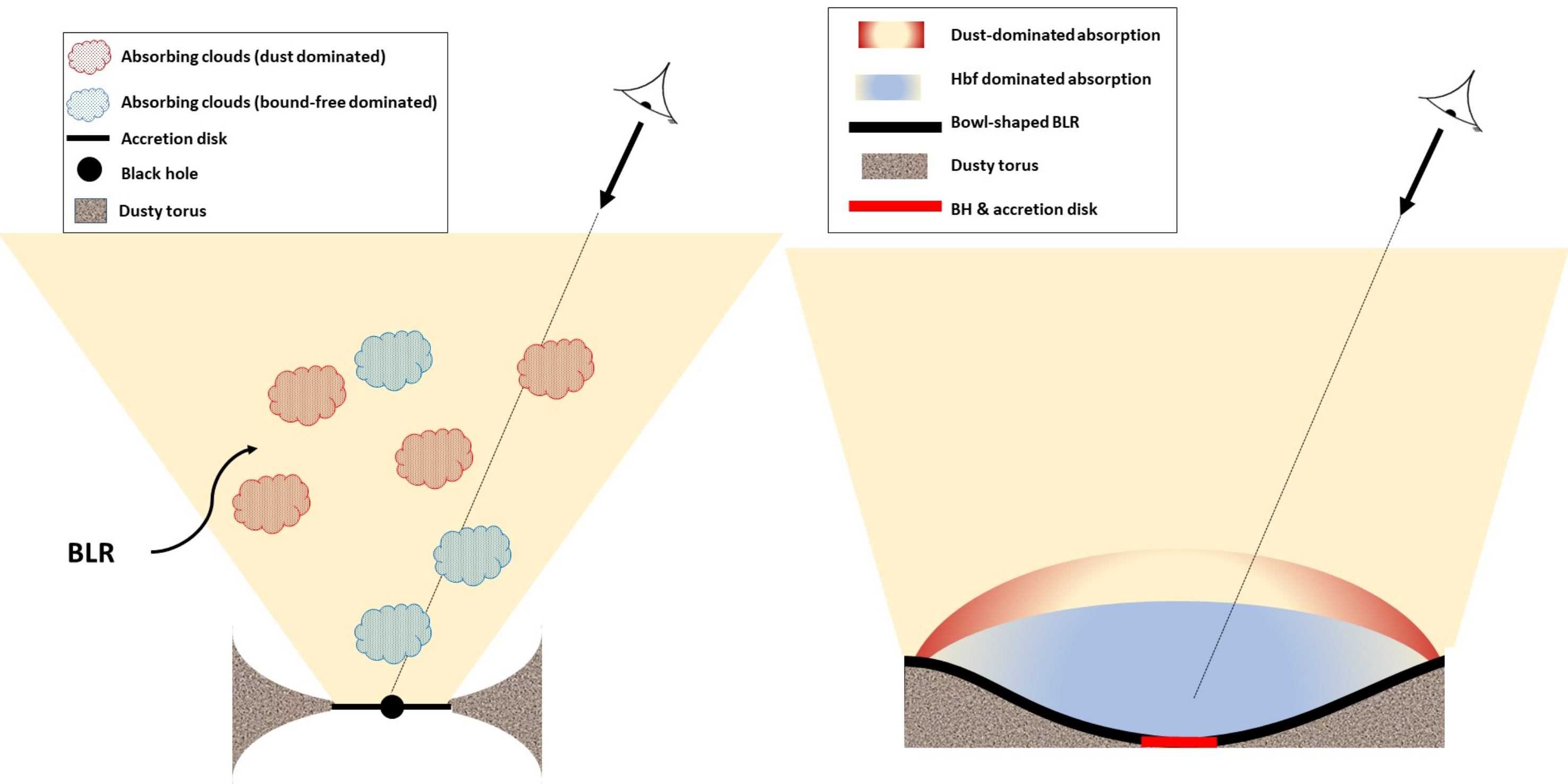}
	\caption{Illustration of differential \Hb/\paa{} absorption. (left panel) by clouds intercepting the line of sight, or (right panel) by an atmosphere above the accretion disk/BLR with different densities and central concentrations of dust/Hbf absorption. The black disk is the accretion disk and flattened BLR.}
	\label{fig:shirnlsresults12}
\end{figure*}

The \paa{} flux is mostly close to or above that expected for \Hb{} flux, as indicated by the fiducial lines.The optical depth at the relevant wavelengths for \Hb{} and \paa{} depends on whether the optical depth down the line of sight is dominated by hydrogen bound-free (Hbf) opacity or reddening.  The sense of this model is illustrated in Figure \ref{fig:shirnlsresults12} (left panel), where the BLR clouds intercept the line of sight to the observer, contributing to the several absorption modes. Cases where FWHM(\Hb)~$>$~FWHM(\paa{}) can arise where the optical spectrum has lower spatial resolution (e.g. the 6.7\AA~ fibres of 6dF), leading to a larger diversity of velocities or alternatively because the optical depth of broad line clouds in \Hb{} is smaller than their optical depth in \paa{}. As mentioned before, we discount the former case; the latter case will occur when the dominant opacity is H bound-free and free-free opacity. Cases where FWHM(\Hb)~$<$~FWHM(\paa) can arise where extinction by dust is the dominant opacity. Clouds responsible for reddening do not have to be the same clouds as those emitting hydrogen line radiation; they just have to be mixed in with them. If the reddening dominates, \paa{} will be seen from closer to the SMBH and that line is broader. If Hbf dominates, \Hb{} will be formed closer to the SMBH and is broader. 

\begin{figure}
	\centering
	\includegraphics[width=1\linewidth]{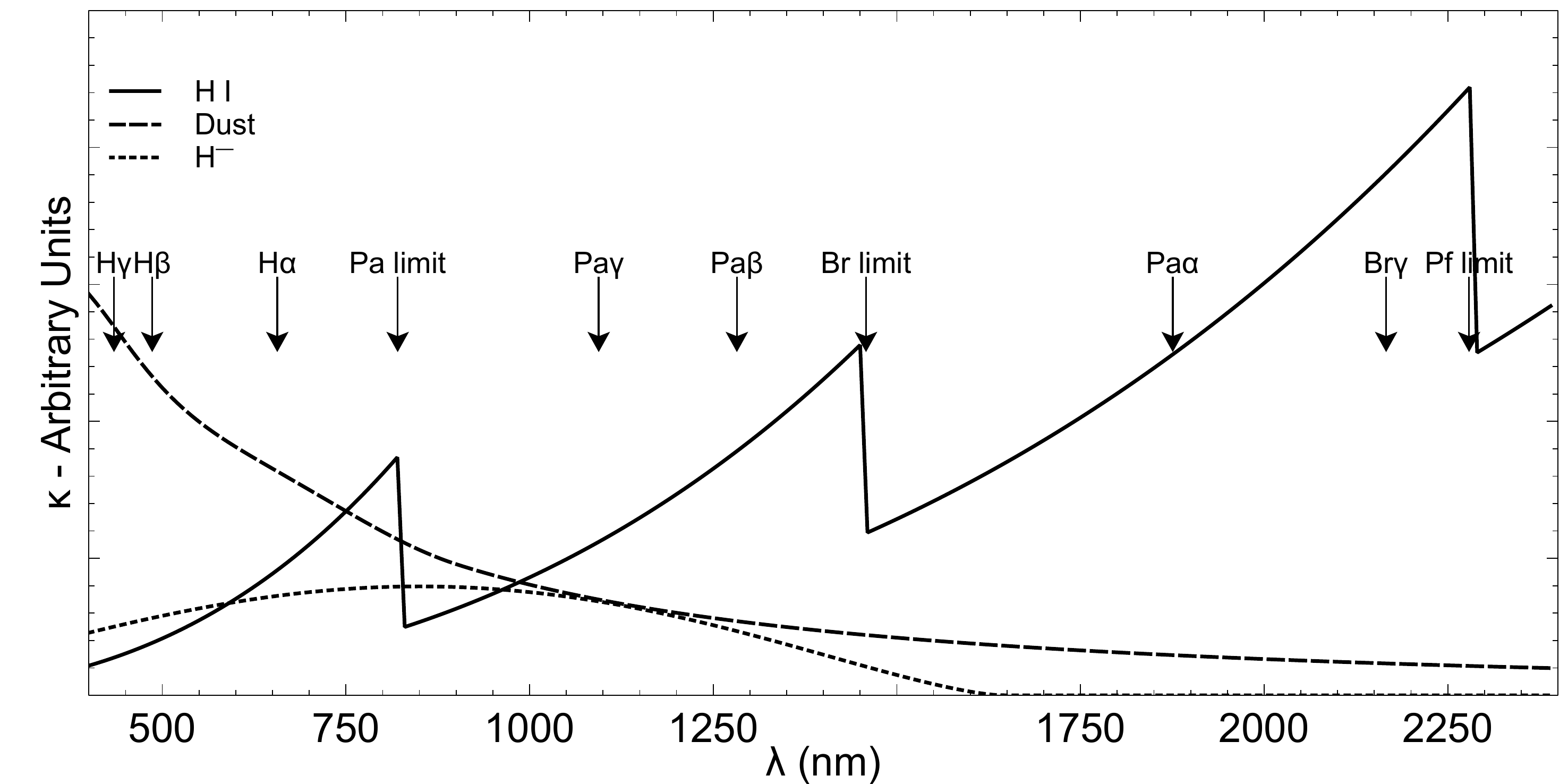}
	\caption{Relative opacity vs. wavelength for contributions from dust, H I bound-free and H\pwr{-} absorption. Each plot is arbitrarily scaled for illustrative purposes. The wavelengths of Balmer, Paschen and Brackett series lines are shown with the series limits pointed out (including the Pfund series). Note the sharp jump in opacity shortward of each limit.}
	\label{fig:shirnlsresults13}
\end{figure}

As an example, we compute the opacity from three sources: reddening caused by dust (using the \cite{Cardelli1989} extinction law), the Hbf opacity (using the ATLAS9 program of \citealp{Kurucz1970}) and H\pwr{-} opacity (from \cite{Mathisen1984a}, following \cite{Wishart1979} and \citealp{Broad1976}). These opacities are plotted in Figure \ref{fig:shirnlsresults13}, and have each been arbitrarily scaled for illustrative purposes, with the relative contributions determined by the particular distribution of clouds.

From this toy model, we can compute a quantitative example of cloud sizes and distances from the SMBH. For an electron density of $n_e~\sim10^8$ cm$^{-3}$, matter density of $\rho ~\sim$ 1.6 $\times$ 10$^{-16}$ and an opacity of $\kappa~\sim$ 1 cm$^2$gm\pwr{-1}, the optical depth $\tau$ is approximately 1 for a path length $x$ = 0.6 $\times$ 10$^{16}$ cm = 400 AU. For a SMBH of 10$^7M_\odot$ and a line half width of 1200 km/s, the distance $r$ from the SMBH is 6000 AU ($\sim$0.03 pc).

In a two dimensional plane parallel model consisting of a disk centered on the SMBH, path lengths are magnified by $\sec \theta$, where $\theta$ is the angle between the line of sight and the axis of the disk. That will affect the statistics of the line width ratios.
\subsection{Differential Absorption by Atmosphere}
Another scenario is to posit an atmosphere above the accretion disk, where the flux/velocity structure is determined by the relative \textit{radial} dominance of dust or Hbf absorption. If we consider the accretion disk as radiating uniformly per unit area and with a Keplerian velocity structure, then the highest velocities (wings of the line profile) are generated towards the centre of the disk (but with low flux) and the highest flux is generated further out (but with lower velocities). The width of the line is then determined by whether the central region is dominated by the particular absorption mode. In the plot of flux and FWHM ratios (Figure \ref{fig:shirnlsresults11} left panel), the numbered regimes have properties as follows:
\begin{enumerate}[label={\arabic*.},leftmargin=*,align=left]
    \item Hbf absorption reduces the total flux of \paa, but dust dominates in the central region.
    \item Dust absorption dominates overall, so \Hb{} flux and widths are less than \paa. 
    \item Hbf absorption dominates overall, so \Hb{} flux and widths are greater than \paa.
    \item Dust absorption reduces the total flux of \Hb, but Hbf absorption dominates in the central region.
\end{enumerate}
The distribution of measurements shows that in the majority of cases, dust extinction is the main contributor to the relative absorption, but with several objects showing increased Hbf absorption in the central region. The sense of this model is illustrated by Figure \ref{fig:shirnlsresults12} (right panel), where the centrally located Hbf absorption is indicated in blue, and the peripheral dust absorption in red. We recognise that measurement uncertainties may distort this picture somewhat.
\section{Conclusions}
\label{sec:Sec6}
We have presented a near-infrared spectroscopic survey (using the SOFI instrument on the ESO-NTT telescope) of narrow-line Seyfert 1 galaxies in the southern hemisphere. These have been sampled from the optical surveys of \cite{Chen2018} and \cite{Schmidt2018}. We have examined the kinematics of the broad-line region, probed by the emission line width of hydrogen (\paa{} and \Hb). We find that a single Lorentzian fit (preferred on theoretical grounds) to the line profiles is as good (if not better) than multi-component Gaussian fits. A lack of measurable NLR emission of \Fe{} and \Htwo{} and thus of hydrogen (overwhelmed by the pole-on view of the BLR light) supports a single Lorentzian (rather than a multiple Gaussian) fit.

The BH mass estimates from the original catalogs were calculated by either the \LBCont{} and FWHM of \Hb{} (\citetalias{Chen2018}) or by luminosity/FWHM of \Ha{} (\citetalias{Schmidt2018}). We recomputed these using the values of FWHM and luminosity of \Hb{} given in the catalogs; and found a relationship slope greater than unity compared to the original values. We ascribed this to contamination by galactic light or confusion with the \NII{} line flux. We also recomputed the BH masses by remeasuring the \Hb{} line using a Lorentzian fit; this again produced a relationship slope greater than unity, compared to the catalog values. However, the comparison of masses computed by the Lorentzian and Gaussian measurements showed a slope close to unity.

We estimated the BH masses using \paa{} and compared them with those from the optical emission line \Hb. This comparison of line width and flux ratios shows deviations from expected luminosity ratios and equality of line widths. We posit a scenario where an intermixture of dusty and dust-free clouds (or alternately a structured atmosphere) differentially absorbs the line radiation of the broad-line emission region, due to dust absorption and hydrogen bound-free absorption.

We also examined the extinction towards the BLR using two sets of line ratios, \Ha/\Hb{} and \paa/\Hb. While these are correlated, the scatter indicates different origins for the optical and infra-red lines.
\section*{Acknowledgements}
Based on observations collected at the European Organisation for Astronomical Research in the Southern Hemisphere under ESO program 0103.B-0504(B). This publication makes use of data products from the Wide-field Infrared Survey Explorer (\textit{WISE}), which is a joint project of the University of California, Los Angeles, and the Jet Propulsion Laboratory/California Institute of Technology, funded by the National Aeronautics and Space Administration (NASA) and from the Two Micron All Sky Survey (2MASS), which is a joint project of the University of Massachusetts and the Infrared Processing and Analysis Center/California Institute of Technology, funded by the NASA and the National Science Foundation. This research has also made use of TOPCAT \citep{Taylor2005} and the VizieR Astronomical Server. 
\section*{Data Availability}
The observational data and the tables in the appendix, as well as the complete set of plots of the Lorentzian emission line fits of \paa{} and \Hb, are available as supplementary material to this paper, and also at CDS via anonymous ftp to \url{cdsarc.u-strasbg.fr} (\url{130.79.128.5}) or via \url{https://cdsarc.unistra.fr/viz-bin/cat/J/MNRAS/???/???}.
\bibliographystyle{mnras}
\bibliography{library} 

\appendix
\section{Tables}
\begin{landscape}
\begin{table}
\setlength\extrarowheight{2pt}
\caption{Observation Log. (Sample)}
\label{tbl:shirnlsobslog}
	\small
\begin{tabular}{llrrrrrrcccr}
\toprule
Object&Alt. ID&\multicolumn{1}{c}{RA$^a$}&\multicolumn{1}{c}{Dec$^a$}&\multicolumn{1}{c}{\textit{K}$^b$}&	\multicolumn{1}{c}{z$^a$}&Lum. Dist.$^c$&\ebv$^d$&ABBA$^e$&Telluric&Obs. Date&S/N$^f$\\ 
&&&&(mag)&&(Mpc)&(mag)&Cycles&Star&(Aug 2019)&\\
\midrule
6dF-J004039.2-371317   &                         & 10.16321  & -37.22133 & 12.30 & 0.0360 & 158            & 0.015  & 1           & HIP005675  & 24       & 26.3 \\
6dF-J020349.0-124717   & 2MASX J02395613-1118128 & 30.95430  & -12.78799 & 13.37 & 0.0525 & 234            & 0.023  & 1           & HIP012055  & 24       & 15.3 \\
6dF-J022815.2-405715   &                         & 37.06350  & -40.95410 & 11.92 & 0.4934 & 2817           & 0.036  & 1           & HIP006806  & 24       & 12.0 \\
6dF-J023005.5-085953   & MRK 1044                & 37.52301  & -8.99811  & 10.83 & 0.0161 & 70             & 0.023  & 1           & HIP012055  & 24       & 7.6  \\
6dF-J041307.1-005017   & 2MASX J04472072-0508138 & 63.27955  & -0.83799  & 12.71 & 0.0400 & 177            & 0.118  & 1           & HIP019830  & 24       & 15.1 \\
6dF-J044040.4-411044   &                         & 70.16812  & -41.17881 & 12.14 & 0.0327 & 144            & 0.021  & 1           & HIP021020  & 24       & 16.3 \\
6dF-J224520.3-465211   &                         & 341.33458 & -46.86983 & 11.92 & 0.2000 & 985            & 0.035  & 1           & HIP112735  & 24       & 5.2  \\
2MASXJ05014863-2253232 &                         & 75.45257  & -22.88979 & 12.32 & 0.0410 & 181            & 0.022  & 1           & HIP023671  & 25       & 17.3 \\
\bottomrule
\multicolumn{12}{l}{\textbf{Notes:}}\\
\multicolumn{12}{l}{\textit{a} RA, Dec and redshift (z) from are from the NASA/IPAC Extragalactic Database (NED).}\\
\multicolumn{12}{l}{\textit{b} \textit{K} apparent magnitude from the 2MASS catalog \citep{Skrutskie2006}.}\\
\multicolumn{12}{l}{\textit{d} Galactic extinction \url{https://irsa.ipac.caltech.edu/applications/DUST/} with the \cite{Schlafly2011} values and \cite{Cardelli1989} extinction law.}\\
\multicolumn{12}{l}{\textit{e} Number of observing cycles (nodding ABBA)}\\
\multicolumn{12}{l}{\textit{f} S/N ratio, see Section \ref{sec:observations}.}\\
\end{tabular}
\end{table}
\begin{table}
\centering
\setlength\extrarowheight{2pt}
\caption{\paa{} and \Hb{} emission line measurements, extinction and \mbh. (Sample)}
\label{tbl:shirnlsmeasure}
\scriptsize
\begin{tabular}{@{}l|cccccc|cccccc@{}}
\toprule
&\multicolumn{6}{c|}{\paa}&\multicolumn{6}{c}{\Hb}\\
Object            & Flux (10\pwr{-15}   & FWHM    & log(L)    & V       & log(\mbh)   &\ebv      & Flux (10\pwr{-15}      & FWHM    & log(L)    & V       & log(\mbh)  &\ebv       \\
&\ecs)&(nm)     & (\es)     &(\kms)        & \msun   &(mag) $\dagger$       &1\ecs)&(nm)     & (\es)     &(\kms)        & \msun   &(mag) $\ddagger$ \\ \midrule
2MASXJ01413249-1528016 & 14.7$\pm$2.8    & 5.5$\pm$0.3  & 41.38$\pm$0.08 & 727.2$\pm$38.6  & 6.58$\pm$0.09 & 0.74$\pm$0.08  & 6.1$\pm$0.7    & 14.0$\pm$1.4 & 41.00$\pm$0.05 & 839.0$\pm$86.4   & 6.52$\pm$0.12 & 0.48$\pm$0.12  \\
2MASXJ05014863-2253232 & 28.7$\pm$3.7    & 3.1$\pm$0.1  & 41.05$\pm$0.06 & 499.7$\pm$16.2  & 6.09$\pm$0.06 & 0.58$\pm$0.05  & 19.4$\pm$1.6   & 11.3$\pm$0.8 & 40.88$\pm$0.04 & 667.3$\pm$46.2   & 6.26$\pm$0.08 & 0.33$\pm$0.08  \\
6dF-J000040.3-054101   & 26.8$\pm$1.9    & 6.0$\pm$0.1  & 41.78$\pm$0.03 & 827.1$\pm$15.2  & 6.89$\pm$0.03 & 0.60$\pm$0.03  & 17.2$\pm$1.3   & 14.0$\pm$0.9 & 41.58$\pm$0.03 & 810.8$\pm$50.1   & 6.77$\pm$0.07 & 0.20$\pm$0.09  \\
6dF-J004039.2-371317   & 35.9$\pm$4.7    & 7.2$\pm$0.2  & 41.03$\pm$0.06 & 1029.8$\pm$30.4 & 6.71$\pm$0.05 & 0.48$\pm$0.05  & 32.7$\pm$1.4   & 23.2$\pm$0.9 & 40.99$\pm$0.02 & 1399.9$\pm$53.0  & 6.96$\pm$0.04 & 0.18$\pm$0.05  \\
6dF-J013335.0-210959   & 7.1$\pm$0.6     & 3.6$\pm$0.1  & 41.50$\pm$0.04 & 499.7$\pm$11.7  & 6.31$\pm$0.04 & 0.34$\pm$0.03  & 9.8$\pm$0.5    & 7.7$\pm$0.3  & 41.64$\pm$0.02 & 364.8$\pm$16.7   & 6.11$\pm$0.05 & 0.42$\pm$0.05  \\
6dF-J013546.4-353915   & 12.8$\pm$0.8    & 4.8$\pm$0.1  & 41.81$\pm$0.03 & 586.3$\pm$10.1  & 6.60$\pm$0.03 & -0.08$\pm$0.03 & 60.6$\pm$3.3   & 18.2$\pm$0.8 & 42.49$\pm$0.02 & 1083.1$\pm$50.3  & 7.47$\pm$0.05 & 0.02$\pm$0.06  \\
6dF-J014222.3-352541   & 5.0$\pm$1.1     & 7.1$\pm$0.4  & 41.05$\pm$0.10 & 1025.2$\pm$56.0 & 6.72$\pm$0.09 & -0.28$\pm$0.07 & 42.4$\pm$1.4   & 17.3$\pm$0.5 & 41.98$\pm$0.01 & 1025.9$\pm$29.1  & 7.17$\pm$0.03 & -0.02$\pm$0.04 \\
\bottomrule
\multicolumn{13}{l}{\textbf{Notes:}}\\
\multicolumn{13}{l}{$\dagger$ \ebv{} from \paa/\Hb.}\\
\multicolumn{13}{l}{$\ddagger$ \ebv{} from \Ha/\Hb.}\\
\multicolumn{13}{l}{\dots{} \Ha{} not measurable.}\\\end{tabular}
\end{table}
\end{landscape}
\bsp
\label{lastpage}
\end{document}